\documentclass{article}
\usepackage{authblk}
\usepackage{fullpage}
\usepackage{graphicx}%
\usepackage{multirow}%
\usepackage{amsmath,amsfonts}%
\usepackage{amsthm}%
\usepackage{mathrsfs}%
\usepackage[title]{appendix}%
\usepackage{xcolor}%
\usepackage{textcomp}%
\usepackage{manyfoot}%
\usepackage{booktabs}%
\usepackage{algorithm}%
\usepackage{algorithmicx}%
\usepackage{algpseudocode}
\usepackage{stfloats}
\usepackage{cuted}

\usepackage{mathtools}
\usepackage{tikz}
\usetikzlibrary{calc,tikzmark}
\usetikzlibrary{positioning}
\usetikzlibrary{circuits.logic.US} %
\usetikzlibrary{decorations.pathreplacing}
\usetikzlibrary{matrix}
\usepackage{nicematrix}
\usetikzlibrary{shapes.geometric}
\usetikzlibrary{fit}
\usetikzlibrary{positioning}
\usetikzlibrary {arrows.meta}
\usetikzlibrary{calc,intersections,through,backgrounds}
\tikzstyle{cm dotted}=[dash pattern=on \pgflinewidth off 2mm ]
\usepackage{caption}
\usepackage{subcaption}
\usepackage{blkarray}
\usepackage{multicol}
\usepackage{multirow}
\usepackage{enumitem}
\usepackage{tabularx, makecell, booktabs}
\usepackage{xcolor}

\AtBeginDocument{%
  \providecommand\BibTeX{{%
    \normalfont B\kern-0.5em{\scshape i\kern-0.25em b}\kern-0.8em\TeX}}}

\title{A quantum model for synchronizing finite state transition systems}

\author[1]{Martin~Lukac}
\author[2]{Khaled~El-Fakih}
\author[3]{Uraz Cengiz T\"urker}

\affil[1]{Hiroshima City University, Hiroshima, Japan, malu@hiroshima-cu.ac.jp}
\affil[2]{Aemerican University of Sharjah, Sharjah, UAE, kelfakih@aus.edu}
\affil[3]{University of Lancaster, Lancaster, UK, u.turker@lancaster.ac.uk}
\date{}
\begin{document}
\maketitle
\begin{abstract}
We propose a quantum model for finding a resetting input sequence (RS) which can take a finite state transition system (FA), to particular state independent of its current state. The complexity of finding such sequences for various types of FA can be NP-Hard or even PSPACE-Complete. To this end, we represent the FA states, inputs, and transition function in quantum space. Accordingly, we propose a model to represent the execution of an input sequence of a particular length $l$ starting form an initial FA state. The model is extended considering the application in superposition of all input sequences of length $l$ to an initial state of the FA. The model is further extended considering the application of all input sequences to all initial states of the FA capturing for every input sequence the collection (ordered list) of states reached by applying the sequence to all states of the FA.
The amplitude amplification algorithm is then used as it combines similar collections of reached states while preserving all input sequences that reach these collections. A Grover search for a reached collection where its elements correspond to the same FA state provides a RS for the FA. Our approach offers a quadratic gain over the exponential complexity of traditional brute-force method, which is the only method that can be applied to a general FA class. 
\end{abstract}






\maketitle

\section{Introduction}
State transition-based systems are an umbrella term to represent systems that change their internal states after receiving some inputs. These systems include, but are not limited to, hardware, software, mathematical, and biological systems. In general, state-transition-based systems must return to a particular initial (or consistent/safe) state to function correctly. This operation is known as the \textit{reset operation}. In essence  a reset operation  defines the steps required to bring the system back to a particular initial or consistent state independent of its current state. Because of its role, resets are indispensable aspects of many systems. Some systems use reset circuits to carry out this operation. In case of lack of reset circuits, a reset operation is done by executing a sequence of inputs called \textit{reset sequence} (RS)
(also named as \textit{synchronizing sequence} or a \textit{reset word}).

In robotics, for example, RSs are crucial for multi-robot manipulation and transportation, where multiple robots must coordinate and recover from faults to ensure the system operates smoothly~\cite{RSinRobots22}.

Automotive control systems also benefit from these sequences, as they help synchronize embedded systems in vehicles, maintaining safety and functionality in critical systems after unexpected failures~\cite{RSinAutomaticControl}. 

In communication systems, RSs ensure the synchronization of protocol handlers after disruptions, enabling uninterrupted data flow and maintaining the integrity of communication networks~\cite{DBLP:conf/pts/LuoDBVG93}. 

RSs are also key in software engineering; namely, in conformance testing of reactive systems where many tests or input words are derived to detect errors or unexpected behaviour of a given implementation. The implementation needs to be reset before running each test~\cite{Chow78, pomeranz95,lee96}.

In bio-computing, a collection of automata work in parallel on different inputs and in order to start a new computation a RS is applied to reset the automata to their initial states. In ~\cite{benson00} for example, the researchers showed that in a controlled environment they ran $3 * 10^{12}$ automata per microliter, performing $6 * 10^{10}$ transitions per second, and in order to run these automata, we need to run reset words. The reader may also refer to other applications of RSs in robotics ~\cite{ananichev1}, matrix theory~\cite{Gerencs18}, and Markov processes~\cite{travers2011exact}.

A Finite Automaton (FA) is a fundamental mathematical computation model used to describe state-transition-based systems' behaviour. An FA consists of finite numbers of states, inputs, and transitions between states, and it can be in exactly one of its states at a given time and can transition from one state to another in response to inputs it receives. Depending on the underlying system, many FA classes exist, each having different properties.  For example, if all system states share a common set of inputs, the system can be represented by a {\em complete FA}; otherwise, one can use a {\em partial FA}. 

Automated RS derivation from FA is a long-standing problem. In 1990, Epsstein used a quadratic algorithm to find an RS from a complete FA~\cite{eppstein}.  Moreover, as the length of the RS is the time required to reset a given system, deriving one of the shortest RS is usually needed. In 1964, Černý conjectured that for every $n$-state complete FA having an RS, there is an RS of length at most $(n-1)^2$~\cite{cerny64,Cerny71}. This conjecture is unproved yet. However, Eppstein also showed that for complete FA, finding the one of the shortest RS is an NP-Hard problem~\cite{eppstein}. They also revealed that checking the existence of an RS of length $l$ from a complete FA is NP-Complete. 

For partial FA, which is the general case, the picture is more pessimistic as the length of the shortest RS is $\Omega(3^{n/3})$ and the length of an RS is bounded by $\Theta(n^2*4^{n/3})$ ~\cite{gazdag}. Therefore, for such systems modelled by a partial FA, checking the existence of an RS is PSPACE-Complete~\cite{martyugin}. Therefore, in the literature, there exists a line of research to investigate efficient algorithms for finding (checking existence, deriving, deriving shortest) RSs ~\cite{eppstein,Roman05, RafiqC03,Roman12,RS15,RS97, volkov2008,volkov2009,Kisielewicz2,TURKER201669,
KarahodaEKTY20,SARAC2021114203,GunicenEY13, Kari21,Shabana22,Kemal2021}.

Quantum computers offer the promise of up to exponential speed-ups over classical computers. Recent advances in quantum computing hardware have resulted in quantum processors with more than 400 qubits~\cite{Patra_2024}, and there are claims that quantum superiority has been achieved on different quantum devices, including superconducting systems ~\cite{Arute2019} and photonic quantum devices ~\cite{zhong2020}. Quantum algorithms like Grover's algorithm ~\cite{grover} and amplitude amplification give polynomial speed-ups over corresponding classical algorithms. Yet, though these algorithms give comparably modest quadratic speed-up, they are widely applicable and thus give speed-ups for a wide range of problems ~\cite{Nielsen10}. Some other algorithms like Shor algorithm~\cite{Shor97} offer exponential speed-up.

Despite extensive efforts in designing classical algorithms for RS generation, the computational complexity of deriving shortest RSs remains a major challenge, especially for partial automata where classical methods are often infeasible due to their exponential time complexity.

While quantum algorithms have been successfully applied to a wide range of search and optimisation problems, their application to reset sequence generation remains unexplored. To the best of our knowledge, no formal method has yet been proposed for constructing quantum circuits that generate reset sequences for finite automata.

In this paper, we introduce a quantum RS generation algorithm. This is the first formal method to generate a quantum circuit to compute an RS from a given (complete/partial) FA.  
A high-level summary of the method is provided in the following section. A simple application example, the rules for transforming a transition function into a permutative one and a proof about the space complexity are provided in the online supplementary material provided in the appendix.

\section{Background}\label{sec:background}

A Finite Automata (FA)  $M = (S,X,\delta )$ has a finite set $S$ of $n$ states, a finite set $X$ of $k$ inputs, and a next-state transition function: $\delta : S \times X \rightarrow S$. If $M$ is at state $p$ (can also be denoted by $s_p$, $p=0...n-1$) and input $x$ (inputs can also be denoted as $x_i$, $i$ $\in$ $0..k-1$) is applied, then $M$ moves to state $q = \delta(p, x)$.




In quantum computing, the information is represented by quantum bits, qubits. A qubit represents the information in a complex Hilbert space by a set of orthonormal basis vectors in a wave equation. For a single qubit represented by a pair of orthonormal basis states $\vert0\rangle = \begin{pmatrix}
    1\\0
\end{pmatrix}$ and $\vert1\rangle = \begin{pmatrix}
    0\\1
\end{pmatrix}$, the corresponding wave equation $\vert\psi\rangle = \alpha\vert 0\rangle + \beta\vert 1\rangle$, with $\alpha, \beta\in\mathbb{C}$ are complex coefficients for each basis state of the wave equation such that $\vert\alpha\vert^2+\vert\beta\vert^2 = 1$.

When using more than one qubit the multi-qubit quantum state is obtained using the Kronecker product $\otimes$. Given two qubits $\vert\psi\rangle$ and $\vert\phi\rangle = \gamma\vert0\rangle + \delta\vert1\rangle$ their joint state obtianed by Kronecker product is shown in eq.~\ref{eq:skron}.
\begin{equation}
    \vert \rho\rangle = \vert\psi\rangle\otimes\vert\phi\rangle = \alpha\gamma\vert00\rangle+\alpha\delta\vert01\rangle+\beta\gamma\vert 10\rangle+\beta\delta\vert11\rangle\\
    \label{eq:skron}
\end{equation}

 Kronecker product can be also applied to matrices in a similar manner as it is applied to quantum states. 
As an example, let $H=\frac{1}{\sqrt{2}}\begin{pmatrix}
    1&1\\1&-1
\end{pmatrix}$
be the Hadamard quantum gate used to put a quantum state in superposition. Let $\vert\rho\rangle=\vert\psi\rangle\otimes\vert\phi\rangle =\vert0\rangle\otimes\vert1\rangle = \begin{pmatrix} 1\\0\end{pmatrix}\otimes\begin{pmatrix} 0\\1\end{pmatrix}=\begin{pmatrix} 0\\1\\0\\0\end{pmatrix}$. Then to apply the Hadamard transform to both qubit, first we need to create a two qubit Hadamard quantum gate using the kronecker product and then multiply it with the two-qubit state $\vert\rho\rangle$ as shown in eq.~\ref{eq:kron}.
\begin{align}
    H\otimes H\vert\rho\rangle &= \frac{1}{2}\begin{pmatrix}
    1&1\\1&-1
\end{pmatrix}\otimes\begin{pmatrix}
    1&1\\1&-1
\end{pmatrix}\begin{pmatrix} 0\\1\\0\\0\end{pmatrix}\nonumber=\frac{1}{2}\begin{pmatrix}
    1&1&1&1\\1&-1&1&-1\\1&1&-1&-1\\
    1&-1&-1&1
\end{pmatrix}\begin{pmatrix} 0\\1\\0\\0\end{pmatrix}\nonumber\\& = \frac{1}{2}\left ( \vert00\rangle-\vert01\rangle+\vert10\rangle-\vert11\rangle\right )\label{eq:kron} 
\end{align}

Each complex coefficient in a quantum state represents the probability $prob(q)$ of obtaining the state $q$ from the superposition under the measurement. For instance, $\vert\alpha\vert^2$ is the probability to measure the state $\vert0\rangle$ . For instance in the state in equation~\ref{eq:kron}, each two qubit state is equiprobable under measurement with probability of observation being $prob(q) = \frac{1}{4}$.

Finally, the tensor (Kronecker) product applied to a same state or the same gate $k$-times can also be written in a shortenned form. For instance $\vert1\rangle^{\otimes^k}=\vert111\ldots 11\rangle$ represents a quantum register of $k$ qubits all in the basis state $1$. Similarly, $H^{\otimes^k}=H\otimes \ldots\otimes H\otimes H$ represents $k$ Hadamard gates multiplied by kronecker product.

The quantum register can be manipulated using unitary operators such as $H$. A particular type of unitary operator is also called a permutative matrix. A \textit{permutative matrix }is a a square binary matrix that has exactly one entry of 1 in each row and each column with all other entries 0. Thus a permutative matrix contains exactly $n$ one's and the rest are zeros for a matrix of size $n\times n$. 


\section{The approach}\label{sec:results}
\subsection{High-level summary}

Our approach has the following steps:

1) \textit{Represent a given FA in quantum space}: We map the states and input symbols of the FA to quantum registers. To simulate the transition function $\delta$ of the FA under an input symbol $x$, we first construct a unitary matrix that encodes $\delta(x)$. If this matrix is not permutative (i.e., not directly applicable in quantum computation), we transform it into a permutative unitary matrix $\Delta_x$ by introducing ancillary (ancilla) registers. These ancilla registers allow the extension of the system’s Hilbert space so that the resulting matrix is reversible and unitary. A global permutative transition operator $\Delta$ is then derived using all $\Delta_x$ matrices.


2) \textit{Simulate an FA transition in quantum space}: 
To simulate a transition from an initial FA state under a given input, we first construct a quantum total state (QTS) by taking the tensor product of: (i) the quantum encoding of the input symbol, (ii) its corresponding ancilla register, (iii) and the quantum state representing the current FA state.

We then apply the operator $\Delta$ to this QTS. The resulting QTS represents: (i) the input symbol, (ii) the reached FA state (as a quantum state), (iii) and any garbage data left in the ancilla registers after the computation.


3) \textit{Simulate the application of an input sequence}: 
We extend the previous step to simulate the effect of an input sequence of length $l$. We apply the operator $\Delta$ iteratively $l$ times, denoted $\Delta^l$, to the tensor product of: (i) the quantum encoding of the full input sequence, (ii) the associated ancilla registers, and (iii) the initial FA state. The final QTS reflects the FA state reached after applying the entire input sequence.


4) \textit{Simulate the application of all input sequences}: 
To simulate the application of all input sequences of length $l$ simultaneously, we initialize the system in a uniform superposition over all possible input sequences of that length. The QTS is constructed as the tensor product of: (i) the superposition of all input sequences, (ii) the corresponding ancilla registers, and (iii) the initial FA state.

Applying $\Delta^l$ in this setting results in a quantum state that contains, for each sequence, the corresponding reached FA state and ancillary outputs.


5) \textit{Simulate the application of all input sequences starting from all FA states}: We extend step (4) by simulating, in superposition, the effect of all input sequences of length $l$ on all FA states. The initial quantum state now encodes all FA states, and the total quantum state (QTS) is formed by the tensor product of this encoding with all input sequences of length $l$ and their ancilla registers. For each sequence, the resulting QTS captures all quantum states reached when that sequence is applied to every FA state.


6) \textit{Use Grover algorithm to search for an intended RS}: The construction in (5) groups together the FA states that evolve into identical quantum states in addition to having the input sequences that reach such identical states. Grover’s algorithm is then used to find such sequences—those that map all FA states to the same final state yielding an RS. For partial FAs, the FA is first completed by adding a “don’t care” state with self-loops for all inputs and for each FA state with an undefined input, a transition under the input is added connecting the state to the don't care state. The added state is included in the quantum search.



\subsection{Representing FA states and inputs and transition function in quantum space}


\textit{Representing the FA states and inputs}: 

Each orthonormal basis $\vert0\rangle$, $\vert1\rangle$,... can be mapped to a state of the FA. Therefore, the $n$ states of an FA can be directly mapped to the $n$ basis vectors represented in a quantum state using $\log_2 n$ qubits. Accordingly, we define: 

\begin{itemize}
\item The $\nu$-qubit register is used to represent the state $\vert\mathcal{S}\rangle =\vert w_{\nu-1}\ldots w_0\rangle$, with $\nu = \log_2n$, and the basis states $\vert \mathcal{S}_0\rangle$, $\vert \mathcal{S}_1\rangle$, ..., $\vert \mathcal{S}_{n-1}\rangle$. 
\end{itemize}

Similarly, for an FA with $k$ inputs, the inputs can be represented in quantum space using $\log_2 k$ qubits. Accordingly, we define:

\begin{itemize}
\item  A $\kappa$-qubit register representing the input $\vert\mathcal{X}\rangle =\vert\xi_{\kappa-1}\ldots \xi_0\rangle$, with $\kappa = \log_2k$, and the basis states $\vert \mathcal{X}_0\rangle$, $\vert \mathcal{X}_1\rangle$, ...,$\vert \mathcal{X}_{k-1}\rangle$
\end{itemize}

\textit{Representing FA transition function}:
The state change of an FA is defined by the state transition function $\delta$. 
To achieve the mapping of $\delta$ from state $s_i$ ($\vert\mathcal{S_{i}}\rangle$) and input $x$ leading to state $s_j$ ($\vert\mathcal{S_{j}}\rangle$), 
we first map $\delta$ to a unitary square binary matrix $\bar{\Delta}_x$ which has exactly one entry equals to $1$ in each row (specifies next-quantum state) and in each column (current quantum state), and $0$ elsewhere. Namely, if $\delta_x(s_i,x)=s_j$, then 1 is placed in the intersection of the column representing $\vert \mathcal{S}_i\rangle$ and the row representing $\vert \mathcal{S}_j\rangle$. The matrix $\bar{\Delta_x}$ must be permutative so that it can be directly mapped to a quantum implementation (required by the implementation limitations of reversible and quantum computing). However, if for certain input $x$, the corresponding unitary matrix that encodes $\delta(x)$ is not permutative, then $\bar{\Delta}_x$ is transformed into a permutative matrix by introducing $\mu$ ancilla qubits register denoted by $\vert\mathcal{A}\rangle$.

In fact, we define the following: 
\begin{itemize}
        \item A $\mu$-qubit register representing ancilla $\vert\mathcal{A}\rangle =\vert a_{\mu-1}\ldots a_0\rangle$ with $\mu = \log_2m$ with basis states $\vert\mathcal{A}_0\rangle\ldots\vert\mathcal{A}_{m-1}\rangle$. 
\end{itemize}

Note, first, for each input $x$ with a non-permutative $\bar{\Delta}_{x_i}$, 
we determine $\mu'$ the ancillae necessary to create the corresponding unitary permutative matrix. The total number of required ancillae over all inputs $x$ is computed as $\mu = \arg\max_{\mu'}\max(\mu')$. Then, considering $\mu$, for every input $x$, FA embedding can be used to derive the corresponding permutative $\Delta_{x}$ using the general approaches for embedding reversible automata or functions ~\cite{DBLP:journals/corr/SoekenWKMD14}.

In our case, the columns and rows of $\Delta_{x}$
are labeled by the basis vectors of the tensor product $\vert\mathcal{A}\mathcal{S}\rangle = \vert\mathcal{A}\rangle\otimes\vert\mathcal{S}\rangle$. 

By construction of a $\Delta_{x_i}$, for each transition
$\delta(s_k, x_i)=s_l$, ${\Delta}_{x_i}[\vert\mathcal{A}_?\mathcal{S}_?
\rangle][\vert\mathcal{A}_?\mathcal{S}_?\rangle]=1$.

Then we define $\Delta$ representing the $\delta$ next-state function under all $x\in X$ as a block-diagonal (permutative) matrix where for each $x$ there is a corresponding $\Delta_x$. Namely, we place in the diagonal of $\Delta$, the individual $\Delta_{x}$ operators ordered in the natural order of the elements in $X$. The first operator will be $\Delta_{x_0}$, then $\Delta_{x_1}$ and so on, as shown in eq.~\ref{eq:u}. All other elements of the matrix are set to zeros.

\color[rgb]{0,0,0}

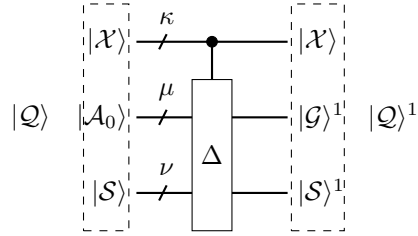
\begin{figure}[h]
\centering
\begin{tikzpicture}

\filldraw[black] (1,1.5) circle (2pt);
\draw[black,thick]  (0,1.5)node[anchor=east,inner xsep=3pt,inner ysep=1pt]{$\vert \mathcal{X}\rangle$} -- (2,1.5)node[anchor=west] {$\vert\mathcal{X}\rangle$};
\draw[black,thick](0.3,1.4) -- (0.4,1.6)node[anchor=south]{$\kappa$};
\draw[black,thick]  (0,0.5)node[anchor=east,inner xsep=3pt,inner ysep=1pt]{$\vert \mathcal{A}_0\rangle$} -- (2,0.5)node[anchor=west] {$\vert\mathcal{G}\rangle^1$};
\draw[black,thick](0.3,0.4) -- (0.4,0.6)node[anchor=south]{$\mu$};
\draw[black,thick]  (0,-0.5)node[anchor=east,inner xsep=3pt,inner ysep=1pt]{$\vert \mathcal{S}\rangle$} -- (2,-0.5)node[anchor=west] {$\vert\mathcal{S}\rangle^1$};
\draw[black,thick](0.3,-0.6) -- (0.4,-0.4)node[anchor=south]{$\nu$};
\node[rectangle,draw,minimum width = 0.5cm, minimum height = 2cm,fill=white] (d1) at (1,0) {$\Delta$};
\draw[black,thick]  (1,1.5)node[anchor=east,inner xsep=3pt,inner ysep=1pt]{} -- (d1.north);
\node[rectangle,draw,minimum width = 0.6cm, minimum height = 3cm,dashed] (r1l) at (-0.4,0.5){};
\node[left of=r1l] (r1ll) {$\vert\mathcal{Q}\rangle$};
\node[rectangle,draw,minimum width = 0.7cm, minimum height = 3cm,dashed] (r2l) at (2.4,0.5){};
\node[right of=r2l] (r1ll) {$\vert\mathcal{Q}\rangle^1$};
\end{tikzpicture}
    \caption{Circuit representing the application of an FA input $x$ encoded by $\vert \mathcal{X}\rangle$ to an FA state encoded by $\vert \mathcal{S}\rangle$ according to the $\Delta$ function. This represents the transition from QTS $\vert\mathcal{Q}\rangle$ to $\vert\mathcal{Q}\rangle^1$}\label{fig:input}
\end{figure}

\begin{figure*}[b!]
     \centering
      \resizebox{0.7\textwidth}{!}{
\begin{tikzpicture}
\def\xz{1};
\def\yz{0};

\def\xo{\xz+4};
\def\yo{\yz};

\def\xt{\xo+4};
\def\yt{\yo};
\def\yi{4}
\def\ydi{0.5}
\def\ya{2}
\def\yda{0.5}
\def\ys{0}

\filldraw[black] (\xt,\yi+2*\ydi) circle (2pt);
\filldraw[black] (\xo,\yi+\ydi) circle (2pt);
\filldraw[black] (\xz,\yi) circle (2pt);
\draw[cm dotted, line width=.25mm,step=0.5]  (\xz+0.2,\yi+.1) -- (\xo-0.1,\yi+\ydi-.1);
\draw[cm dotted, line width=.25mm,step=0.5]  (\xo+0.2,\yi+\ydi+.1) -- (\xt-0.1,\yi+2*\ydi-.1);
\draw[cm dotted, line width=.25mm,step=0.5]  (\xz+0.6,\ya-\yda) -- (\xo-.55,\ya-\yda);
\draw[cm dotted, line width=.25mm,step=0.5]  (\xo+0.6,\ya-\yda) -- (\xt-.55,\ya-\yda);
\draw[cm dotted, line width=.25mm,step=0.5]  (\xz-0.75,\ya+.1) -- (0.25,\ya+\yda-.1);
\draw[cm dotted, line width=.25mm,step=0.01]  (\xz-0.75,\ya+\yda+.1) -- (0.25,\ya+2*\yda-.1);

\draw[black,thick]  (\xz-1,5)node[anchor=east,inner xsep=3pt,inner ysep=1pt]{$\vert \mathcal{X}^l_j\rangle$} -- (\xt+1,\yi+2*\ydi)node[anchor=west,inner xsep=3pt,inner ysep=1pt]{$\vert \mathcal{X}^l_j\rangle$};
\draw[black,thick](\xz-0.8,\yi+2*\ydi-0.1) -- (\xz-0.7,\yi+2*\ydi+0.1)node[anchor=south,yshift=-0.1cm]{$\kappa$};
\draw[black,thick]  (\xz-1,4.5)node[anchor=east,inner xsep=3pt,inner ysep=1pt]{$\vert \mathcal{X}^{l-1}_h\rangle$} -- (\xt+1,\yi+\ydi)node[anchor=west,inner xsep=3pt,inner ysep=1pt]{$\vert \mathcal{X}^{l-1}_h\rangle$};
\draw[black,thick](\xz-0.8,\yi+\ydi-0.1) -- (\xz-0.7,\yi+\ydi+0.1)node[anchor=south,yshift=-0.1cm]{$\kappa$};
\draw[black,thick]  (\xz-1,4)node[anchor=east,inner xsep=3pt,inner ysep=1pt]{$\vert \mathcal{X}^1_i\rangle$} -- (\xt+1,\yi)node[anchor=west,inner xsep=3pt,inner ysep=1pt]{$\vert \mathcal{X}^1_i\rangle$};
\draw[black,thick](\xz-0.8,\yi-0.1) -- (\xz-0.7,\yi+0.1)node[anchor=south,yshift=-0.1cm]{$\kappa$};

\draw[black,thick]  (\xz-1,\ya+\yda*2)node[anchor=east,inner xsep=3pt,inner ysep=1pt]{$\vert \mathcal{A}_0\rangle^l$} -- (\xz+.55,\ya+2*\yda)node[anchor=west,inner xsep=3pt,inner ysep=1pt]{$\vert \mathcal{A}_0\rangle^l$};
\draw[black,thick](\xz-0.8,\ya+\yda*2-0.1) -- (\xz-0.7,\ya+2*\yda+0.1)node[anchor=south,yshift=-0.1cm]{$\mu$};
\draw[black,thick]  (\xz-1,\ya+\yda)node[anchor=east,inner xsep=3pt,inner ysep=1pt]{$\vert \mathcal{A}_0\rangle^{l-1}$} -- (\xz+.55,\ya+\yda)node[anchor=west,inner xsep=3pt,inner ysep=1pt]{$\vert \mathcal{A}_0 \rangle^{l-1}$};
\draw[black,thick](\xz-0.8,\ya+\yda-0.1) -- (\xz-0.7,\ya+\yda+0.1)node[anchor=south,yshift=-0.1cm]{$\mu$};
\draw[black,thick]  (\xz-1,\ya)node[anchor=east,inner xsep=3pt,inner ysep=1pt]{$\vert \mathcal{A}_0\rangle^1$} -- (\xz+.55,\ya)node[anchor=west,inner xsep=3pt,inner ysep=1pt]{$\vert \mathcal{G}\rangle^1$};
\draw[black,thick](\xz-0.8,\ya-0.1) -- (\xz-0.7,\ya+0.1)node[anchor=south,yshift=-0.1cm]{$\mu$};
\draw[black,thick]  (\xz-1,\ys+1)node[anchor=east,inner xsep=3pt,inner ysep=1pt]{$\vert \mathcal{S}\rangle$} -- (\xz+.55,\ys+1)node[anchor=west,inner xsep=3pt,inner ysep=1pt]{$\vert \mathcal{S}\rangle^1$};
\draw[black,thick](\xz-0.8,\ys+1-0.1) -- (\xz-0.7,\ys+1+0.1)node[anchor=south,yshift=-0.1cm]{$\nu$};

\draw[black,thick]  (\xo-.5,\ya+2*\yda)node[anchor=east,inner xsep=3pt,inner ysep=1pt]{$\vert \mathcal{A}_0\rangle^l$}  -- (\xo+.55,\ya+2*\yda)node[anchor=west,inner xsep=3pt,inner ysep=1pt]{$\vert \mathcal{A}_0\rangle^l$};
\draw[black,thick] (\xo-.5,\ya+\yda)node[anchor=east,inner xsep=3pt,inner ysep=1pt]{$\vert \mathcal{A}_0\rangle^{l-1}$}  -- (\xo+.55,\ya+\yda)node[anchor=west,inner xsep=3pt,inner ysep=1pt]{$\vert \mathcal{G}\rangle^{l-1}$};
\draw[black,thick] (\xo-.5,\ya)node[anchor=east,inner xsep=3pt,inner ysep=1pt]{$\vert \mathcal{G}\rangle^1$}  -- (\xo+.55,\ya)node[anchor=west,inner xsep=3pt,inner ysep=1pt]{$\vert \mathcal{G}\rangle^1$};
\draw[black,thick]  (\xo-.5,\ys+1)node[anchor=east,inner xsep=3pt,inner ysep=1pt]{$\vert \mathcal{S}\rangle^{l-2}$}  -- (\xo+.55,\ys+1)node[anchor=west,inner xsep=3pt,inner ysep=1pt]{$\vert \mathcal{S}\rangle^{l-1}$};

\draw[black,thick]  (\xt-.5,\ya+2*\yda)node[anchor=east,inner xsep=3pt,inner ysep=1pt]{$\vert \mathcal{A}_0\rangle^l$} -- (\xt+1,\ya+2*\yda)node[anchor=west,inner xsep=3pt,inner ysep=1pt]{$\vert \mathcal{G}\rangle^l$};
\draw[black,thick] (\xt-.5,\ya+\yda)node[anchor=east,inner xsep=3pt,inner ysep=1pt]{$\vert \mathcal{G}\rangle^{l-1}$} -- (\xt+1,\ya+\yda)node[anchor=west,inner xsep=3pt,inner ysep=1pt]{$\vert \mathcal{G}\rangle^{l-1}$};
\draw[black,thick] (\xt-.5,\ya)node[anchor=east,inner xsep=3pt,inner ysep=1pt]{$\vert \mathcal{G}\rangle^1$} -- (\xt+1,\ya)node[anchor=west,inner xsep=3pt,inner ysep=1pt]{$\vert \mathcal{G}\rangle^1$};
\draw[black,thick]  (\xt-.5,\ys+1)node[anchor=east,inner xsep=3pt,inner ysep=1pt]{$\vert \mathcal{S}\rangle^{l-1}$} -- (\xt+1,\ys+1)node[anchor=west,inner xsep=3pt,inner ysep=1pt]{$\vert \mathcal{S}\rangle^l$};

\node[rectangle,draw,minimum width = 0.5cm, minimum height = 3cm,fill=white] (d1) at (\xz,2) {$\Delta$};
\node[rectangle,draw,minimum width = 0.5cm, minimum height = 3cm,fill=white] (d2) at (\xo,2) {$\Delta$};
\node[rectangle,draw,minimum width = 0.5cm, minimum height = 3cm,fill=white] (d3) at (\xt,2) {$\Delta$};

\node[rectangle,draw,minimum width = 0.8cm, minimum height = 5cm,dashed] (r1l) at (-0.5,3){};
\node[left of=r1l] (r1ll) {$\vert\mathcal{Q}\rangle$};
\node[rectangle,draw,minimum width = 0.8cm, minimum height = 5cm,dashed] (rl) at (\xt+1.45,3){};
\node[right of=rl] (rll) {$\vert\mathcal{Q}\rangle^l$};

\draw[black,thick]  (\xz,\yi)node[anchor=east,inner xsep=3pt,inner ysep=1pt]{} -- (d1.north);
\draw[black,thick]  (\xo,\yi+.5)node[anchor=east,inner xsep=3pt,inner ysep=1pt]{} -- (d2.north);
\draw[black,thick]  (\xt,\yi+1)node[anchor=east,inner xsep=3pt,inner ysep=1pt]{} -- (d3.north);
\end{tikzpicture}}

    \caption{Circuit representing the $l-$ times application of $\Delta$ to a sequence of $l$ inputs encoded by $\vert\mathcal{X}^l_j\rangle\cdots\vert\mathcal{X}^1_i\rangle$ starting from an FA state encoded by $\mathcal{S}\rangle$. This represents the evolution, according to $\Delta$, from QTS $\vert\mathcal{Q}\rangle$ to $\vert\mathcal{Q}\rangle^l$ leading to the FA state encoded by $\mathcal{S}\rangle^l$.}\label{fig:seq} 
\end{figure*}
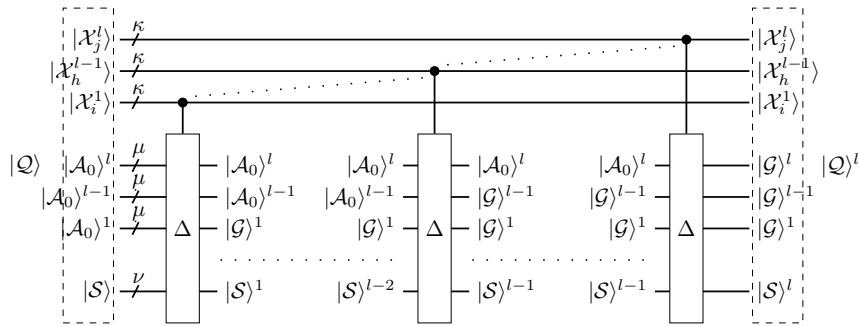


\tikzset{highlight/.style={rectangle,
                           draw=blue,
                           rounded corners = 0.5 mm, 
                           inner sep=0.5pt,
                           fit=#1}}

\begin{equation}
    \Delta = \begin{pNiceMatrix}[margin=3pt,create-medium-nodes,first-row,first-col]
&\vert \mathcal{X}_0\mathcal{E}\rangle&\vert  \mathcal{X}_1\mathcal{E}\rangle&\Cdots&\vert \mathcal{X}_{k-1}\mathcal{E}\rangle\\
\vert \mathcal{X}_0\mathcal{E}\rangle&\Delta_{x_0}&0&\Cdots&0\\\vert \mathcal{X}_1\mathcal{E}\rangle&0&\Ddots&\Ddots&\Vdots\\\Vdots&\Vdots&\Ddots&&0\\\vert \mathcal{X}_{k-1}\mathcal{E}\rangle&0&\Cdots&0&\Delta_{x_{k-1}}\\
\CodeAfter
\begin{tikzpicture}[name suffix = -medium]
\node [highlight=(1-1)(1-1)(1-1)(1-1)] {} ; 
\node [highlight=(4-4)(4-4)(4-4)(4-4)] {} ;
\end{tikzpicture}
\end{pNiceMatrix}
\label{eq:u}
\end{equation}



To conveniently represent the combinations of the basis states of $\vert\mathcal{A}\rangle$ and $\vert\mathcal{S}\rangle$ we use $\vert\mathcal{E}\rangle =\vert\mathcal{A}\rangle\otimes\vert\mathcal{S}\rangle=\vert\mathcal{A}_0\mathcal{S}_0\rangle+\ldots+\vert\mathcal{A}_{m-1}\mathcal{S}_{n-1}\rangle$ to annotate the rows and columns of the matrix $\Delta$. Therefore, each term $\vert\mathcal{X}_i\mathcal{E}\rangle$ can be expanded as a vector $\vert\mathcal{X}_i\mathcal{A}_0\mathcal{S}_0\rangle \ldots\vert\mathcal{X}_i\mathcal{A}_{m-1}\mathcal{S}_{n-1}\rangle$. 

We use $\vert\mathcal{Q}\rangle$, referred to as Quantum Total State (QTS), to represent the tensor product of the three defined registers $\vert\mathcal{X}\rangle$, 
 $\vert\mathcal{A}\rangle$, and 
 $\vert\mathcal{S}\rangle$, respectively, as shown in eq.~\ref{eq:genstate}.

\begin{equation}
    \vert\mathcal{Q}\rangle = \vert\mathcal{X}\rangle\vert\mathcal{A}\rangle\vert\mathcal{S}\rangle 
\label{eq:genstate} 
\end{equation}

Each square ($\Delta_{x_i}$) in eq.~\ref{eq:u} shows one specific operator mapping $\mathcal{A}\times \mathcal{S}\rightarrow \mathcal{A}\times \mathcal{S}$ for a specific input value $x_i$. 
Thus, for an input encoded in $\vert\mathcal{X}\rangle$, using the $\Delta$ operator, we transform $\vert\mathcal{Q}\rangle$ from eq.~\ref{eq:genstate} to the next-quantum state $\vert\mathcal{Q}\rangle^1$ as shown in eq.~\ref{eq:genstate1}. The ancilla $\vert\mathcal{A}\rangle$ is always set as $\vert\mathcal{A}\rangle=\vert\mathcal{A}_0\rangle =\vert0\rangle^{\otimes \mu}$ before the $\Delta$ operation is applied. Observe that the application of $\Delta$ transforms the ancillae $\vert\mathcal{A}\rangle$ into garbage $\vert\mathcal{G}\rangle^1$. The garbage indicates that it should be discarded because it is used and its value is unknown. We note that both the used garbage and the resulting quantum state $\vert\mathcal{S}\rangle^1$ are indexed by a superscript indicating the application of one input. The circuit representing the application an FA input encoded by $\vert\mathcal{X}\rangle$ to an FA state encoded by $\vert\mathcal{S}\rangle$ according to $\Delta$ leading to the FA state encoded by $\vert\mathcal{S}\rangle^1$ is 
shown in Figure~\ref{fig:input}.


\begin{equation}
    \Delta \vert\mathcal{Q}\rangle = \vert\mathcal{Q}\rangle^1 = \Delta\vert\mathcal{X} \rangle   \vert\mathcal{A}_0\rangle\vert\mathcal{S}\rangle =
    \vert\mathcal{X} \rangle \vert\mathcal{G}\rangle^1\vert\mathcal{S}\rangle^1
\label{eq:genstate1} 
\end{equation}

The number of qubits to encode inputs (states) is obtained directly from $log_2$ of the total number of input values (FA states). 




\color[rgb]{0,0,0}

\begin{table}[bht]
\centering
\caption{\label{tab:M1} An FA Example}
\begin{tabular}{llll}
\toprule
& Input  &   &    \\
State  & $x_0$  & $x_1$  &    \\
\midrule
$s_0$ & $s_1$ & $s_0$ &    \\
$s_1$ & $s_2$ & $s_2$ &    \\
$s_2$ & $s_0$ & $s_2$ &    \\
\bottomrule
\end{tabular}
\end{table}

\textit{Working example}: As an application example, consider the FA given in Table ~\ref{tab:M1}. The FA has three states $S = \{s_0,s_1,s_2\}$, two inputs $X = \{x_0,x_1\}$, and four transitions. 
 First, we map the states $s_0 \rightarrow \vert0\rangle = \begin{pmatrix}1\\0\\0\end{pmatrix}=\vert\mathcal{S}_0\rangle$, $s_1\rightarrow \vert1\rangle = \begin{pmatrix}0\\1\\0\end{pmatrix} =\vert\mathcal{S}_1\rangle$  and $s_2\rightarrow \vert2\rangle = \begin{pmatrix}0\\0\\1\end{pmatrix} =\vert\mathcal{S}_2\rangle$. We decide to encode the  states by qutrits (quantum bits with three basis states).  The input values are encoded each by a single qubit $\vert\mathcal{X}_0\rangle = \vert0\rangle=\begin{pmatrix}1\\0\end{pmatrix}$ and $\vert\mathcal{X}_1\rangle = \vert1\rangle=\begin{pmatrix}0\\1\end{pmatrix}$.

Considering the input $x_1$, we transform the state transition function (Table~\ref{tab:M1}) into a the matrix $\bar{\Delta}_{x_1}$ 
shown in eq.
~\ref{eq:nonunit}. 
 \begin{equation}
     \bar{\Delta}_{x_1}=\begin{pNiceMatrix}[first-row,first-col,create-medium-nodes]
     &\vert \mathcal{S}_0\rangle&\vert \mathcal{S}_1\rangle&\vert \mathcal{S}_2\rangle\\
         \vert \mathcal{S}_0\rangle&1&0&0\\
         \vert \mathcal{S}_1\rangle&0&0&0\\
         \vert \mathcal{S}_2\rangle&0&1&1\\
     \end{pNiceMatrix}
     \label{eq:nonunit}
 \end{equation}
where if $\delta(s_i,x_1) = s_j$  then $1$ is placed at intersection of column $\vert \mathcal{S}_i\rangle$ and row $\vert \mathcal{S}_j\rangle$  denoted as $\bar{\Delta}_{x_1}[\vert\mathcal{S}_j\rangle][\vert\mathcal{S}_i\rangle]=1$. 

Note that  as $\delta(s_0,x_1) = s_0$, we have in eq.~\ref{eq:nonunit} 
$\bar{\Delta}_{x_1}[\vert\mathcal{S}_0\rangle][\vert\mathcal{S}_0\rangle]=1$. 
In addition, observe that as $\delta(s_1,x_1) =\delta(s_2,x_1) = s_3$, the matrix $\bar{\Delta}_{x_1}$ is non-permutative. This is shown by $\bar{\Delta}_{x_1}[\vert\mathcal{S}_2\rangle][\vert\mathcal{S}_1\rangle]=1$ and $\bar{\Delta}_{x_1}[\vert\mathcal{S}_2\rangle][\vert\mathcal{S}_2\rangle]=1$. For the input $x_2$, we observe that $\bar{\Delta}_{x_2}$ is permutative as $\delta(s_0,x_0) \neq \delta(s_1,x_0) \neq \delta(s_2,x_0)$. Thus, In order to transform $\bar{\Delta}_{x_1}$ into a permutative $\Delta_{x_1}$, we need one ancilla qubit because $m=2$; in this FA there are at maximum two state transitions leading to the same successor state. And for the input $x_2$. The ancillae qubit is thus represented by a wave equation with two basis states: $\vert\mathcal{A}\rangle = \frac{1}{\sqrt{2}}(\vert\mathcal{A}_0\rangle+\vert\mathcal{A}_1\rangle)$ with 
basis vectors $\vert \mathcal{A}_0\rangle=\begin{pmatrix}
    1\\0
\end{pmatrix}$ and $\vert \mathcal{A}_1\rangle=\begin{pmatrix}
    0\\1
\end{pmatrix}$. Then using tensor product we create $\vert\mathcal{A}\mathcal{S}\rangle = \vert\mathcal{A}\rangle\otimes\vert\mathcal{S}\rangle = \frac{1}{\sqrt{6}}(\vert\mathcal{A}_0\mathcal{S}_0\rangle+\vert\mathcal{A}_0\mathcal{S}_1\rangle+\vert\mathcal{A}_0\mathcal{S}_2\rangle+\vert\mathcal{A}_1\mathcal{S}_0\rangle+\vert\mathcal{A}_1\mathcal{S}_1\rangle+\vert\mathcal{A}_1\mathcal{S}_2\rangle)$. The unitary matrix $\Delta_{x_1}$ will have the rows and columns labeled by the basis vectors of $\vert\mathcal{A}\mathcal{S}\rangle$. To construct $\Delta_{x_1}$, for each state transition $\delta(s_k,x_1) = s_l$ we place a 1 in column labeled by $\vert\mathcal{A}_0\mathcal{S}_k\rangle$ and any empty row that contains $\vert\mathcal{S}_l\rangle$ in the label. Using this rule we map $\delta(s_0,x_1) = s_0$ to ${\Delta}_{x_1}[\vert\mathcal{A}_0\mathcal{S}_0\rangle][\vert\mathcal{A}_0\mathcal{S}_0\rangle]=1$, $\delta(s_1,x_1) = s_2$ to ${\Delta}_{x_1}[\vert\mathcal{A}_0\mathcal{S}_2\rangle][\vert\mathcal{A}_0\mathcal{S}_1\rangle]=1$ and $\delta(s_2,x_1) = s_2$ to ${\Delta}_{x_1}[\vert\mathcal{A}_1\mathcal{S}_2\rangle][\vert\mathcal{A}_0\mathcal{S}_2\rangle]=1$.  For all the columns that contain $\vert\mathcal{A}_1\rangle$ put a 1 in the remaining empty rows. An example of such a constructed unitary matrix is shown in eq.~\ref{eq:deltax1}. 

\tikzset{highlightr/.style={rectangle,
                           draw=red,
                           rounded corners = 0.5 mm, 
                           inner sep=0.5pt,
                           fit=#1},highlightb/.style={rectangle,
                           draw=blue,
                           rounded corners = 0.5 mm, 
                           inner sep=0.5pt,
                           fit=#1}}
\begin{equation}\resizebox{0.9\linewidth}{!}{$
    \Delta_{x_1} =\begin{pNiceMatrix}[first-row,first-col,create-medium-nodes]
    &\vert \mathcal{A}_0 \mathcal{S}_0\rangle&\vert \mathcal{A}_0\mathcal{S}_1\rangle&\vert \mathcal{A}_0\mathcal{S}_2\rangle&\vert \mathcal{A}_1 \mathcal{S}_0\rangle&\vert \mathcal{A}_1\mathcal{S}_1\rangle&\vert \mathcal{A}_1\mathcal{S}_2\rangle\\
        \vert \mathcal{A}_0 \mathcal{S}_0\rangle&1&0&0&0&0&0\\\vert \mathcal{A}_0\mathcal{S}_1\rangle&0&0&0&1&0&0\\\vert \mathcal{A}_0\mathcal{S}_2\rangle&0&1&0&0&0&0\\\vert \mathcal{A}_1 \mathcal{S}_0\rangle&0&0&0&0&1&0\\\vert \mathcal{A}_1\mathcal{S}_1\rangle&0&0&0&0&0&1\\\vert \mathcal{A}_1\mathcal{S}_2\rangle&0&0&1&0&0&0
\CodeAfter
\begin{tikzpicture}[name suffix = -medium]
\node [highlightr=(1-1)(6-1)(6-3)(3-1)] {} ; 
\end{tikzpicture}     
    \end{pNiceMatrix}
    \label{eq:deltax1}
$}\end{equation}

For the input $x_0$, as $\bar{\Delta}_{x_0}$ is permutative, there is no need to embed $\bar{\Delta}_{x_0}$ with additional ancillae. Thus, we when constructing $\Delta_{x_0}$, we consider the ancilla qubit used for $\Delta_{x_1}$, and accordingly, we obtain $\Delta_{x_0}$ shown in equation ~\ref{eq:deltax0} constructed using the same rules as $\Delta_{x_1}$.

\begin{equation}
    \Delta_{x_0} =\begin{pNiceMatrix}[first-row,first-col,create-medium-nodes]
    &\vert \mathcal{A}_0 \mathcal{S}_0\rangle&\vert \mathcal{A}_0\mathcal{S}_1\rangle&\vert \mathcal{A}_0\mathcal{S}_2\rangle&\vert \mathcal{A}_1 \mathcal{S}_0\rangle&\vert \mathcal{A}_1\mathcal{S}_1\rangle&\vert \mathcal{A}_1\mathcal{S}_2\rangle\\
        \vert \mathcal{A}_0 \mathcal{S}_0\rangle&0&0&1&0&0&0\\\vert \mathcal{A}_0\mathcal{S}_1\rangle&1&0&0&0&0&0\\\vert \mathcal{A}_0\mathcal{S}_2\rangle&0&1&0&0&0&0\\\vert \mathcal{A}_1 \mathcal{S}_0\rangle&0&0&0&0&1&0\\\vert \mathcal{A}_1\mathcal{S}_1\rangle&0&0&0&0&0&1\\\vert \mathcal{A}_1\mathcal{S}_2\rangle&0&0&0&1&0&0
\CodeAfter
\begin{tikzpicture}[name suffix = -medium]
\node [highlightr=(1-1)(6-1)(6-3)(3-1)] {} ; 
\end{tikzpicture}        
    \end{pNiceMatrix}
    \label{eq:deltax0}
\end{equation}
The obtained $\Delta$ contains both matrices $\Delta_{x_0}$ and $\Delta_{x_1}$ as block matrices on the diagonal as shown in eq.~\ref{eq:exdelta}. To represent clearly matrix $\Delta$ we used $\vert\mathcal{E}\rangle =\vert\mathcal{A}\rangle\otimes\vert\mathcal{S}\rangle =\vert\mathcal{A}_0\mathcal{S}_0\rangle+\vert\mathcal{A}_0\mathcal{S}_1\rangle+\vert\mathcal{A}_0\mathcal{S}_2\rangle+\vert\mathcal{A}_1\mathcal{S}_0\rangle+\vert\mathcal{A}_1\mathcal{S}_1\rangle+\vert\mathcal{A}_1\mathcal{S}_2\rangle$ to indicate all combinations of the ancillae and the FA states.

\begin{equation}
\resizebox{0.5\linewidth}{!}{$
    \Delta =\begin{pNiceMatrix}[first-row,first-col,create-medium-nodes]
&&\vert \mathcal{X}_0\rangle\vert\mathcal{E}\rangle&&&\vert \mathcal{X}_1\rangle\vert\mathcal{E}\rangle&\\
&\Block{3-3}<\Large>{\Delta_{x_0}}&&&0&\Cdots&0\\
\vert \mathcal{X}_0\rangle\vert\mathcal{E}\rangle&&&&\Vdots&&\\
&&&&&&\Vdots\\
&&&&0&\Cdots&0\\
&0&\Cdots&0&\Block{3-3}<\Large>{\Delta_{\mathcal{X}_1}}&\\
\vert \mathcal{X}_1\rangle\vert\mathcal{E}\rangle&\\
&\Vdots&&\Vdots&&&\\
&0&\Cdots&0&&&
    \end{pNiceMatrix}$}
    \label{eq:exdelta}
\end{equation}

To simulate the application of $x_1$ to the FA starting from $s_1$, we use the input $\vert\mathcal{X}_1\rangle = \vert1\rangle$, with $\vert\mathcal{S}\rangle = \vert \mathcal{S}_1\rangle = \vert1\rangle$ and $\vert\mathcal{A}\rangle = \vert\mathcal{A}_0\rangle =\vert0\rangle$, and apply $\Delta$ to the QTS $\vert\mathcal{Q}\rangle = \vert 100\rangle$ as shown in eq.~\ref{eq:exgenstate}. The obtained QTS $\vert\mathcal{S}\rangle^1=\vert0\rangle=\vert S_0\rangle$ and $\vert\mathcal{G}\rangle^1=\vert1\rangle=\vert\mathcal{A}_1\rangle$.

\begin{multline}
    \Delta \vert\mathcal{Q}\rangle = \vert\mathcal{Q}\rangle^1 =\Delta \vert \mathcal{X}_1\mathcal{A}_0\mathcal{S}_1\rangle\\ = \Delta \vert 101\rangle=    \Delta\begin{pmatrix}0\\1\end{pmatrix}\otimes\begin{pmatrix}1\\0\end{pmatrix}\otimes\begin{pmatrix}0\\1\\0\end{pmatrix}=\Delta \begin{pmatrix}0\\1\end{pmatrix}\otimes\begin{pmatrix}0\\1\\0\\0\\0\\0\end{pmatrix}=\Delta \begin{pmatrix}0\\0\\0\\0\\0\\0\\0\\1\\0\\0\\0\\0\end{pmatrix}
    =\begin{pmatrix}0\\0\\0\\0\\0\\0\\0\\0\\1\\0\\0\\0\end{pmatrix}=\begin{pmatrix}0\\1\end{pmatrix}\otimes\begin{pmatrix}0\\0\\1\\0\\0\\0\end{pmatrix}=\begin{pmatrix}0\\1\end{pmatrix}\otimes\begin{pmatrix}1\\0\end{pmatrix}\otimes\begin{pmatrix}0\\0\\1\end{pmatrix}\\=\vert1\rangle \vert02\rangle=
    \vert\mathcal{X}_1\rangle \vert\mathcal{A}_0\rangle^1\vert\mathcal{S}_1\rangle^1=\vert\mathcal{X}_1\rangle \vert\mathcal{G}\rangle\vert\mathcal{S}_2\rangle
\label{eq:exgenstate} 
\end{multline}

If we apply the sequence $x^2$ starting from the initial FA state $s_0$; i.e. $\vert S\rangle = \vert\mathcal{S}_0\rangle$, then $\vert\mathcal{G}\rangle^1\vert\mathcal{S}\rangle^1=\vert0\rangle\vert\mathcal{S}_1\rangle=\vert0\rangle\vert1\rangle$ and $\vert\mathcal{S}\rangle^2=\vert\mathcal{S}_2\rangle=\vert2\rangle$ and $\vert\mathcal{G}\rangle^2=\vert0\rangle$.

\color[rgb]{0,0,0}

\subsection{Executing an input sequence starting from a quantum state}
\label{sec:QFAi}

We use $x^l = x_j.....x_i$  to denote an input sequence of length $l$; $i,j \in \{0...k-1\}$. 
We map the inputs of $x^l$ to 
$\vert\mathcal{X}^l_j\rangle\ldots\vert\mathcal{X}^1_i\rangle$, where the superscripts are used to denote the first to the $l$-th inputs of the sequence.  
For each input $\vert\mathcal{X}_i^r\rangle$ where $r=1\ldots l$ and $i\in 0\ldots k-1$, we define an $\mu$-qubit register to represent ancillae $\vert \mathcal{A}_0\rangle=\vert0\rangle^{\otimes \mu}$. We create $l$ ancillae $\vert\mathcal{A}_0\rangle^l\ldots\vert\mathcal{A}_0\rangle^1$ for the $x_j....x_i$ inputs of $x^l$.  The quantum state register $\vert \mathcal{S}\rangle$ contains $\log_2n$ qubits so that it can represent FAs with up to $n$ states by its basis states.  Accordingly, we define the QTS $\vert\mathcal{Q}\rangle$ as the Kronecker product shown in eq.~\ref{eq:qsatatepg}
\begin{equation}
    \vert\mathcal{Q}\rangle = \vert \mathcal{X}^l_j\ldots\mathcal{X}^1_i\rangle\vert\mathcal{A}_0\rangle^l\ldots\vert\mathcal{A}_0\rangle^1\vert\mathcal{S}\rangle
    \label{eq:qsatatepg}
\end{equation}

The application of the $l$-consecutive inputs of $\vert\mathcal{X}^l_j\rangle\cdots\vert\mathcal{X}^1_i\rangle$ starting from QTS $\vert\mathcal{Q}\rangle$ according to the $\Delta$ function leads to QTS $\vert\mathcal{Q}\rangle^l$ as shown in eq.~\ref{eq:evolutions3}. We denote by  $\vert\mathcal{A}_0\rangle^r$ and $\vert\mathcal{G}\rangle^r$, $1\leq r\leq l$, the ancilla and their corresponding garbage registers obtained by the $r-$times application of $\Delta$ (depicted as $\Delta^r$). Similarly, we denote the resulting (reached) quantum state by $\vert\mathcal{S}\rangle^r$. 

\begin{equation}
\Delta^l\vert\mathcal{Q}\rangle=\vert\mathcal{Q}\rangle^l =\Delta(\ldots(\Delta\vert \mathcal{X}^l_j\rangle\ldots\vert\mathcal{X}^1_i\rangle\vert\mathcal{A}_0\rangle^l\ldots\vert\mathcal{A}_0\rangle^1\vert\mathcal{S}\rangle))= \vert \mathcal{X}^l_j\rangle\ldots\vert\mathcal{X}^1_i\rangle\vert \mathcal{G}\rangle^l\ldots\vert\mathcal{G}\rangle^1\vert\mathcal{S}\rangle^l
    \label{eq:evolutions3}
\end{equation}

The circuit representing the application of an input sequence $\vert\mathcal{X}^l_j\rangle\ldots\vert\mathcal{X}^1_i\rangle$ to an FA state encoded by $\vert\mathcal{S}\rangle$ 
 leading to the FA state encoded by $\vert\mathcal{S}\rangle^l$ is shown in Figure~\ref{fig:seq}. Namely, this denotes the $l-$times application of $\Delta$ starting from QTS $\vert\mathcal{Q}\rangle$ and leading to QTS $\vert\mathcal{Q}\rangle^l$. 

 \color[rgb]{0,0,0}
\textit{Working example}: Consider the input sequence ${x}^2= x_0 x_1$ of length $l=2$.
We map the sequence ${x}^2$ onto $\vert\mathcal{X}^2_1\rangle\vert\mathcal{X}^1_0\rangle=\vert10\rangle$.
We require two ancillae, $\vert\mathcal{A}_0\rangle^2\mathcal{A}_0\rangle^1=\vert00\rangle$ one for each application of $\Delta$ and
one qutrit to represent the starting machine state as $\vert\mathcal{S}\rangle$. 
Accordingly, we create the initial QTS shown in eq.~\ref{eq:qts0}.
\begin{equation}
\vert\mathcal{Q}\rangle = \vert\mathcal{X}_1^2\mathcal{X}_0^1\rangle\vert\mathcal{A}_0\rangle^2\mathcal{A}_0\rangle^1\vert\mathcal{S}\rangle
    \label{eq:qts0}
\end{equation}

Performing a state change, starting from the QTS in eq.~\ref{eq:qts0} and using the input sequence $x^2$ is shown in eq.~\ref{eq:evolutions2}. 

\begin{equation}
\Delta^2\vert \mathcal{Q}\rangle=\vert\mathcal{Q}\rangle^2=\Delta\Delta\vert \mathcal{X}^2_1\rangle\vert\mathcal{X}^1_0\rangle\vert \mathcal{A}_0\rangle^2\mathcal{A}_0\rangle^1\mathcal{S}\rangle=\Delta\vert \mathcal{X}^2_1\rangle\vert\mathcal{X}^1_0\rangle\vert \mathcal{A}_0\rangle^2\vert\mathcal{G}\rangle^1\vert\mathcal{S}\rangle^1=\vert \mathcal{X}^2_1\rangle\vert\mathcal{X}^1_0\rangle\vert \mathcal{G}\rangle^2\vert\mathcal{G}\rangle^1\vert\mathcal{S}\rangle^2
    \label{eq:evolutions2}
\end{equation}

If we apply the sequence $x^2$ starting from the initial FA state $s_0$; i.e. $\vert S\rangle = \vert\mathcal{S}_0\rangle$, then $\vert\mathcal{G}\rangle^1\vert\mathcal{S}\rangle^1=\vert0\rangle\vert\mathcal{S}_1\rangle=\vert0\rangle\vert1\rangle$ and $\vert\mathcal{S}\rangle^2=\vert\mathcal{S}_2\rangle=\vert2\rangle$ and $\vert\mathcal{G}\rangle^2=\vert0\rangle$.

\color[rgb]{0,0,0}

\subsection{Executing all possible input sequences starting from a single quantum state}

Here we generalize equation ~\ref{eq:qsatatepg} considering the application of all input sequences of length $l$ to an initial QTS encompassing an initial FA state encoded by $\vert\mathcal{S}\rangle$. 

We recall that the inputs, ancillae, and the quantum states are defined on $\kappa=\log_2 k$, $\mu=\log_2 m$ and $\nu =\log_2 n$ qubits respectively. 
Let's start by a sequence length of $l$, with each input of the sequence $\vert\mathcal{X}_j^l\rangle...\vert\mathcal{X}_i^1\rangle$ being encoded on $\kappa$ qubits encoding $k$ input values. There will be all together $l\kappa$ qubits used to encode an input sequence of $l$ inputs. We create a superposition of all (the $k^l$) input sequences of length $l$ and with $k$ possible input values defined as in eq.~\ref{eq:super}. 
\begin{multline}
\vert\prescript{l}{k}{\mathcal{X}}\rangle=H^{\otimes^{l\kappa}}\vert\mathcal{X}_0^l\rangle\otimes\ldots\otimes\vert\mathcal{X}^1_0\rangle\\=H^{\otimes^{\kappa}}\vert\mathcal{X}_0^l\rangle\otimes\ldots\otimes H^{\otimes^{\kappa}}\vert\mathcal{X}^1_0\rangle=\underbrace{H^{\otimes^{\kappa}}\vert0\rangle^{\otimes^{\kappa}}\otimes\ldots \otimes H^{\otimes^{\kappa}}\vert0\rangle^{\otimes^{\kappa}}}_{l}
\\=\underbrace{H\vert0\rangle\otimes\ldots\otimes H\vert0\rangle}_{H^{\otimes^\kappa}\mathcal{X}^l_0}\otimes\ldots\otimes \underbrace{H\vert0\rangle\otimes\ldots\otimes H\vert0\rangle}_{H^{\otimes^\kappa}\mathcal{X}^1_0}=\underbrace{\frac{1}{\sqrt{k^l}}(\vert0\rangle+\vert1\rangle)\otimes (\vert0\rangle+\vert1\rangle)\otimes\ldots\otimes(\vert0\rangle+\vert1\rangle)}_{kl}\\=\frac{1}{\sqrt{k^l}}\left (\underbrace{\underbrace{\vert0\ldots0\rangle}_{\mathcal{X}^l_0}\ldots\underbrace{\vert0\ldots0\rangle}_{\mathcal{X}^1_0}}_l+\ldots+\underbrace{\underbrace{\vert1\ldots1\rangle}_{\mathcal{X}_{k-1}^l}\ldots\underbrace{\vert1\ldots1\rangle}_{\mathcal{X}_{k-1}^1}}_l \right)=\frac{1}{\sqrt{k^l}}\left (\vert\mathcal{X}_0^l\rangle\ldots\vert\mathcal{X}^1_0\rangle+\ldots+\vert\mathcal{X}_{k-1}^l\rangle\ldots\vert\mathcal{X}^1_{k-1}\rangle \right)
\label{eq:super}
\end{multline}

Combining the superposed input sequences with the ancillae and the state registers, we derive the QTS $\vert\bar{\mathcal{Q}}\rangle=\vert\prescript{l}{k}{\mathcal{X}}\rangle\vert \mathcal{A}_0\rangle^l\ldots \vert \mathcal{A}_0\rangle^1\vert \mathcal{S}\rangle$. 
The application of all input sequences of length $l$ to $\vert\bar{\mathcal{Q}}\rangle$ is shown in equation.~\ref{eq:ginitq3}. 

\begin{multline}
\Delta^l\vert\bar{\mathcal{Q}}\rangle =\vert\bar{\mathcal{Q}}\rangle^l =\Delta^l\vert\prescript{l}{k}{\mathcal{X}}\rangle\vert \mathcal{A}_0\rangle^l\ldots \vert \mathcal{A}_0\rangle^1\vert \mathcal{S}\rangle\\=\Delta^l\frac{1}{\sqrt{{k^l}}}\left (\vert\mathcal{X}^l_0\ldots \mathcal{X}^1_0\rangle\vert \mathcal{A}_0\rangle^l\ldots \vert\mathcal{A}_0\rangle^1\vert \mathcal{S}\rangle
+\ldots+\vert\mathcal{X}^l_{k-1}\ldots \mathcal{X}^1_{k-1}\rangle\vert \mathcal{A}_0\rangle^l\ldots \vert\mathcal{A}_0\rangle^1\vert \mathcal{S}\rangle\right ) 
\\=\frac{1}{\sqrt{k^{l}}}\left (\vert\mathcal{X}^l_0\ldots \mathcal{X}^1_0\rangle\vert \mathcal{G}\rangle^{l,0}\ldots \vert\mathcal{G}\rangle^{1,0}\vert \mathcal{S}\rangle^{l,0}+\ldots+\vert\mathcal{X}^l_{k-1}\ldots \mathcal{X}^1_{k-1}\rangle\vert \mathcal{G}\rangle^{l,k^l-1}\ldots \vert\mathcal{G}\rangle^{1,k^l-1}\vert \mathcal{S}\rangle^{l,k^l-1}\right )
    \label{eq:ginitq3}
\end{multline}

Let $\sigma$ denote the $0... k^r-1$ possible input sequences, over the FA input alphabet, of length $r=1...l$. 
Each input sequence in the superposition of input sequences creates a superposition of the resulting garbage ancillae qubits. We denote by $\vert\mathcal{G}\rangle^{r,\sigma}\ldots\vert\mathcal{G}\rangle^{1,\sigma}$ the garbage qubits resulting from the application of the $\sigma^\text{th}$ input sequence of length $r$ to state $\vert\mathcal{S}\rangle$. Similarly, we use $\vert\mathcal{S}\rangle^{r,\sigma}$ to denote the quantum state reached after the application of the $\sigma^\text{th}$ input sequence of length $r$.

The circuit representing the whole process of putting inputs in superposition, creating all input sequences of length $l$, applying all these input sequences to the FA state encoded by $\vert\mathcal{S}\rangle$ reaching the collection FA states encoded by $\vert\mathcal{S}\rangle^l$ is shown in Figure ~\ref{fig:eQFA1}.

Each quantum state $\vert\mathcal{G}\rangle^{r}....\vert\mathcal{G}\rangle^1\vert\mathcal{S}\rangle^{r} = \sum_{\sigma=0}^{k^r-1}\vert\mathcal{G}\rangle^{r,\sigma}....\vert\mathcal{G}\rangle^{1,\sigma}\vert\mathcal{S}\rangle^{r,\sigma}$ contains  
the FA states encoded by $\vert\mathcal{S}\rangle^{r}$ reached by the application of all the input sequences of length $r$ to the FA state encoded by $\vert\mathcal{S}\rangle$. Figure~\ref{fig:eQFA1} however represents the succession of the QTS evolution in a step by step manner. For instance, $\vert\mathcal{G}\rangle^{1}\vert\mathcal{S}\rangle^{1}$ $=\sum_{\sigma=0}^{k-1}\vert\mathcal{G}\rangle^{1,\sigma}\vert\mathcal{S}\rangle^{1,\sigma}$, represents the states $\vert\mathcal{S}\rangle^1$ reached by the application of the input sequences (denoted by $\sigma=0..k^1-1$) of length $r=1$ to the initial QTS encoded by $\vert\mathcal{S}\rangle$ and the related garbage encoded in $\vert\mathcal{G}\rangle^1$.



Observe that in the circuit of Figure~\ref{fig:eQFA1} each box marked \tikz{\draw[] node[rectangle,draw,black]  at (0,-0.3){H};}  indicates $H^{\otimes^\kappa}$ applied to the qubits encoding each input such as $\vert\mathcal{X}_0^1\rangle =\vert0\ldots0\rangle$.

\begin{figure*}[bht]
    \centering
    \resizebox{\textwidth}{!}{
\begin{tikzpicture}

\draw[black,thick](-1.6,5.4) -- (-1.4,5.6)node[anchor=south,yshift=-0.1cm]{$\kappa$};
\draw[black,thick](-1.6,4.9) -- (-1.4,5.1)node[anchor=south,yshift=-0.1cm]{$\kappa$};
\draw[black,thick](-1.6,3.9) -- (-1.4,4.1)node[anchor=south,yshift=-0.1cm]{$\kappa$};
\draw[black,thick](-1.6,2.9) -- (-1.4,3.1)node[anchor=south,yshift=-0.1cm]{$\mu$};
\draw[black,thick](-1.6,2.4) -- (-1.4,2.6)node[anchor=south,yshift=-0.1cm]{$\mu$};
\draw[black,thick](-1.6,1.4) -- (-1.4,1.6)node[anchor=south,yshift=-0.1cm]{$\mu$};
\draw[black,thick](-1.6,0.9) -- (-1.4,1.1)node[anchor=south,yshift=-0.1cm]{$\nu$};
\draw[black,thick]  (-2,5.5)node[anchor=east,inner xsep=3pt,inner ysep=1pt]{$\vert\mathcal{X}^l_0\rangle$} -- (9.7,5.5)node[anchor=west,inner xsep=3pt,inner ysep=1pt]{$\vert\mathcal{X}^l_0\rangle$}; 
\draw[black,thick]  (-2,4)node[anchor=east,inner xsep=3pt,inner ysep=1pt]{$\vert \mathcal{X}^1_0\rangle$} -- (9.7,4)node[anchor=west,inner xsep=3pt,inner ysep=1pt]{$\vert\mathcal{X}^1_0\rangle$}; 
\draw[black,thick]  (-2,5)node[anchor=east,inner xsep=3pt,inner ysep=1pt]{$\vert\mathcal{X}^{l-1}_0\rangle$} -- (9.7,5)node[anchor=west,inner xsep=3pt,inner ysep=1pt]{$\vert\mathcal{X}^{l-1}_0\rangle$}; 

\draw[black,thick]  (-2,3)node[anchor=east,inner xsep=3pt,inner ysep=1pt]{$\vert \mathcal{A}_0\rangle^l$} -- (1.7,3)node[anchor=west,inner xsep=3pt,inner ysep=1pt]{$\vert \mathcal{A}_0\rangle^l$};
\draw[black,thick]  (-2,2.5)node[anchor=east,inner xsep=3pt,inner ysep=1pt]{$\vert \mathcal{A}_0\rangle^{l-1}$} -- (1.7,2.5)node[anchor=west,inner xsep=3pt,inner ysep=1pt]{$\vert \mathcal{A}_0\rangle^{l-1}$};
\draw[black,thick]  (-2,1.5)node[anchor=east,inner xsep=3pt,inner ysep=1pt]{$\vert \mathcal{A}_0\rangle^1$} -- (1.7,1.5)node[anchor=west,inner xsep=3pt,inner ysep=1pt]{$\vert \mathcal{G}\rangle^1$};
\draw[black,thick]  (-2,1)node[anchor=east,inner xsep=3pt,inner ysep=1pt]{$\vert \mathcal{S}\rangle$} -- (1.7,1)node[anchor=west,inner xsep=3pt,inner ysep=1pt]{$\vert \mathcal{S}\rangle^1$};

\draw[black,thick]  (4.3,3)node[anchor=east,inner xsep=3pt,inner ysep=1pt]{$\vert \mathcal{A}_0\rangle^l$} --(5.7,3)node[anchor=west,inner xsep=3pt,inner ysep=1pt]{$\vert \mathcal{A}_0\rangle^l$};
\draw[black,thick]  (4.3,2.5)node[anchor=east,inner xsep=3pt,inner ysep=1pt]{$\vert \mathcal{A}_0\rangle^{l-1}$} -- (5.7,2.5)node[anchor=west,inner xsep=3pt,inner ysep=1pt]{$\vert \mathcal{G}\rangle^{l-1}$};
\draw[black,thick]  (4.3,1.5)node[anchor=east,inner xsep=3pt,inner ysep=1pt]{$\vert \mathcal{G}\rangle^1$} -- (5.7,1.5)node[anchor=west,inner xsep=3pt,inner ysep=1pt]{$\vert \mathcal{G}\rangle^1$};
\draw[black,thick]  (4.3,1)node[anchor=east,inner xsep=3pt,inner ysep=1pt]{$\vert \mathcal{S}\rangle^{l-2}$} -- (5.7,1)node[anchor=west,inner xsep=3pt,inner ysep=1pt]{$\vert \mathcal{S}\rangle^{l-1}$};

\draw[black,thick]  (8.3,3)node[anchor=east,inner xsep=3pt,inner ysep=1pt]{$\vert \mathcal{A}_0\rangle^l$} -- (9.7,3)node[anchor=west,inner xsep=3pt,inner ysep=1pt]{$\vert \mathcal{G}\rangle^l$};
\draw[black,thick]  (8.3,2.5)node[anchor=east,inner xsep=3pt,inner ysep=1pt]{$\vert \mathcal{G}\rangle^{l-1}$} -- (9.7,2.5)node[anchor=west,inner xsep=3pt,inner ysep=1pt]{$\vert \mathcal{G}\rangle^{l-1}$};
\draw[black,thick]  (8.3,1.5)node[anchor=east,inner xsep=3pt,inner ysep=1pt]{$\vert \mathcal{G}\rangle^1$} -- (9.7,1.5)node[anchor=west,inner xsep=3pt,inner ysep=1pt]{$\vert \mathcal{G}\rangle^1$};
\draw[black,thick]  (8.3,1)node[anchor=east,inner xsep=3pt,inner ysep=1pt]{$\vert \mathcal{S}\rangle^{l-1}$} -- (9.7,1)node[anchor=west,inner xsep=3pt,inner ysep=1pt]{$\vert \mathcal{S}\rangle^l$};

\node[rectangle,draw,fill=white] (h1) at (-0.5,5.5){H};
\node[rectangle,draw,fill=white] (h3) at (-0.5,5){H};
\node[rectangle,draw,fill=white] (h2) at (-0.5,4){H};
\node[rectangle,draw,minimum width = 0.5cm, minimum height = 2.5cm,fill=white] (d1) at (1,2) {$\Delta$};
\node[rectangle,draw,minimum width = 0.5cm, minimum height = 2.5cm,fill=white] (d2) at (9,2) {$\Delta$};
\node[rectangle,draw,minimum width = 0.5cm, minimum height = 2.5cm,fill=white] (d3) at (5,2) {$\Delta$};
\filldraw[black] (9,5.5) circle (2pt);
\filldraw[black] (5,5) circle (2pt);
\filldraw[black] (1,4) circle (2pt);
\draw[black,dotted]  (1.2,4.1) -- (4.9,4.9);
\draw[black,dotted,thick]  (2.2,2) -- (3.8,2);
\draw[black,thick]  (1,4)node[anchor=east,inner xsep=3pt,inner ysep=1pt]{} -- (d1.north);
\draw[black,thick]  (9,5.5)node[anchor=east,inner xsep=3pt,inner ysep=1pt]{} -- (d2.north);
\draw[black,thick]  (5,5)node[anchor=east,inner xsep=3pt,inner ysep=1pt]{} -- (d3.north);

\node[rectangle,draw,minimum width = 1.3cm, minimum height = 5.5cm,dashed] (r1l) at (-2.55,3.2){};
\node[left of=r1l] (r1ll) {$\vert\mathcal{Q}\rangle$};
\node[rectangle,draw,minimum width = 0.4cm, minimum height = 5.5cm,dashed] (rl) at (0.2,3.2){};
\node[above of=rl,yshift=2.1cm] (rll) {$\vert\bar{\mathcal{Q}}\rangle$};
\node[rectangle,draw,minimum width = 1.2cm, minimum height = 5.5cm,dashed] (rfl) at (10.3,3.2){};
\node[above of=rfl,yshift=2.1cm] (rfll) {$\vert\bar{\mathcal{Q}}\rangle^l$};

\draw [decorate,decoration={brace,amplitude=5pt,mirror,raise=4ex}] (1.9,0.7) -- (1.9,1.8){} node[midway,xshift=1cm] (fl1) {};

\node[] (fll1) at (0,0)  {$\vert\mathcal{G}\rangle^1\vert\mathcal{S}\rangle^1 = \sum_{\sigma=0}^{k-1}\vert\mathcal{G}\rangle^{1,\sigma}\vert\mathcal{S}\rangle^{1,\sigma}$};
\draw[black, ->] (fll1.east) -- ++(0.8,0) -- +(0,0.7) -- (fl1.west);

\draw [decorate,decoration={brace,amplitude=5pt,mirror,raise=4ex}] (6.1,0.7) -- (6.1,2.8){} node[midway,xshift=1cm] (fl2) {};

\node[] (fll2) at (5,6.3)  {$\vert\mathcal{G}\rangle^{l-1}\ldots\vert\mathcal{G}\rangle^1\vert\mathcal{S}\rangle^{l-1} = \sum_{\sigma=0}^{k^l-2}\vert\mathcal{G}\rangle^{l-1,\sigma}\ldots\vert\mathcal{G}\rangle^{1,\sigma}\vert\mathcal{S}\rangle^{l-1,\sigma}$};
\draw[black, ->] (fll2.south) -- ++(2.1,-2.5) -- +(0,-1.5) -- (fl2.west);

\draw [decorate,decoration={brace,amplitude=5pt,mirror,raise=4ex}] (10.3,3.6) -- (10.3,5.9){} node[midway,xshift=1.4cm]  {$\vert\prescript{l}{k}{\mathcal{X}}\rangle$};

\draw [decorate,decoration={brace,amplitude=5pt,mirror,raise=4ex}] (10.3,0.7) -- (10.3,3.3){} node[midway,xshift=2.7cm,align=center] (deco)  {};
\node[] (fll3) at (8,0)  {$\vert\mathcal{G}\rangle^{l}\ldots\vert\mathcal{G}\rangle^1\vert\mathcal{S}\rangle^{l} = \sum_{\sigma=0}^{k^l-1}\vert\mathcal{G}\rangle^{l,\sigma}\ldots\vert\mathcal{G}\rangle^{1,\sigma}\vert\mathcal{S}\rangle^{l,\sigma}$};
\draw[black, ->] (fll3.east) -- ++(0.5,0) -- ++(0,1.5) -- +(-0.8,0.4);
\end{tikzpicture}
}
    \caption{Circuit representing the applications of all input sequences of length $l$ in supposition; namely, applying in order the first, second, .., $l-th$ inputs of these sequences starting from a single FA state $\vert\mathcal{S}\rangle$ leading the to the FA states encoded by $\vert\mathcal{S}\rangle^l$. 
    }
    \label{fig:eQFA1}
\end{figure*}
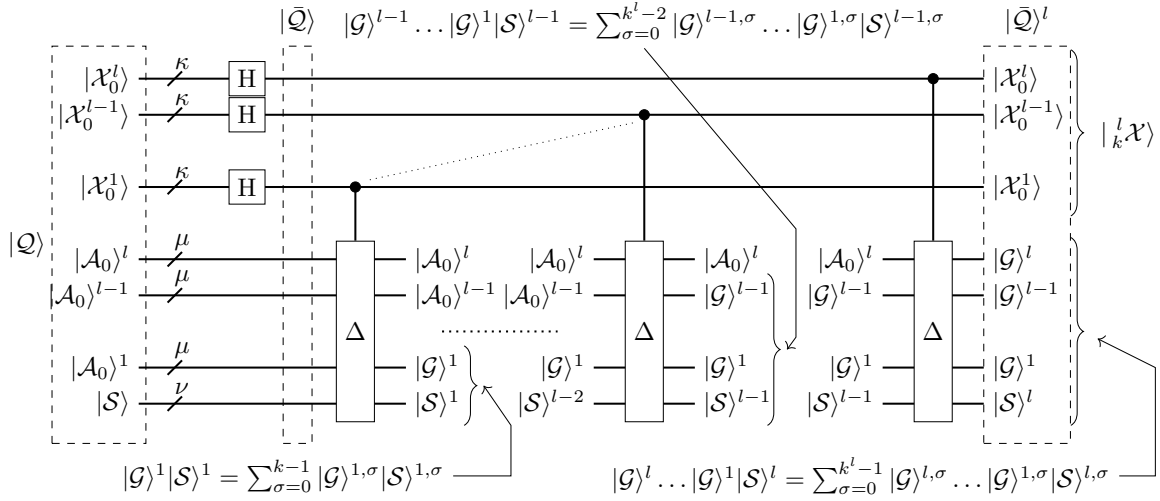


\color[rgb]{0,0,0}
\textit{Working example}: Considering the initial QTS $\vert\mathcal{Q}\rangle = \vert\mathcal{X}_0^2\mathcal{X}_1^1\rangle\vert\mathcal{A}_0\rangle^2\vert\mathcal{A}_0\rangle^1\mathcal{S}\rangle = \vert1100\mathcal{S}\rangle$, we create $\vert\bar{\mathcal{Q}}\rangle$ by setting the input qubits to $0$ and then apply one hadamard quantum gate for each input qubit. The input superposition creation is shown in eq.~\ref{eq:exqbar},
\begin{multline}
\vert\prescript{2}{2}{\mathcal{X}}\rangle=H^{\otimes 2}\vert\mathcal{X}_0^2\mathcal{X}_0^1\rangle=H\vert\mathcal{X}_0^2\rangle\otimes H\vert\mathcal{X}_0^1\rangle=H\vert0\rangle\otimes H\vert0\rangle\\= \frac{1}{\sqrt{2^2}}\left (\vert0\rangle+ \vert1\rangle\right )\left\vert0\rangle+\vert1\rangle\right )= \frac{1}{\sqrt{2^2}}\left (\vert00\rangle+ \vert01\rangle+\vert10\rangle+\vert11\rangle\right )\\= \frac{1}{\sqrt{2^2}}\left (\vert\mathcal{X}^2_0\mathcal{X}^1_0\rangle+ \vert\mathcal{X}^2_0\mathcal{X}^1_1\rangle+\vert\mathcal{X}^2_1\mathcal{X}^1_0\rangle+\vert\mathcal{X}^2_1\mathcal{X}^1_1\rangle\right )
    \label{eq:exqbar}
\end{multline}
and the resulting QTS $\vert\bar{\mathcal{Q}}\rangle$ is shown in eq.~\ref{eq:exqbar1}.
\begin{multline}
    \vert\bar{\mathcal{Q}}\rangle = \vert\prescript{2}{2}{\mathcal{X}}\rangle\vert \mathcal{A}_0\rangle^2\vert\mathcal{A}_0\rangle^1\vert\mathcal{S}\rangle=\frac{1}{\sqrt{2^2}}\left (\vert\mathcal{X}^2_0\mathcal{X}^1_0\rangle\vert\mathcal{A}_0\rangle^2\vert\mathcal{A}_0\rangle^1\vert\mathcal{S}\rangle+ \vert\mathcal{X}^2_0\mathcal{X}^1_1\rangle\vert\mathcal{A}_0\rangle^2\vert\mathcal{A}_0\rangle^1\vert\mathcal{S}\rangle\right .\\\left .+\vert\mathcal{X}^2_1\mathcal{X}^1_0\rangle\vert\mathcal{A}_0\rangle^2\vert\mathcal{A}_0\rangle^1\vert\mathcal{S}\rangle+\vert\mathcal{X}^2_1\mathcal{X}^1_1\rangle\vert\mathcal{A}_0\rangle^2\vert\mathcal{A}_0\rangle^1\vert\mathcal{S}\rangle\right )
    \label{eq:exqbar1}
\end{multline}
Then evolve the initial QTS using all input sequences of length $l=2$. The result is shown in eq.~\ref{eq:init26}.
\begin{multline}
\Delta^2\vert\bar{\mathcal{Q}}\rangle=\vert\bar{\mathcal{Q}}\rangle^2=\Delta^2\vert\prescript{2}{2}{\mathcal{X}}\rangle\vert \mathcal{A}_0\rangle^2\vert\mathcal{A}_0\rangle^1\vert \mathcal{S}\rangle \\= \Delta^2\frac{1}{\sqrt{2^2}}\left (\vert\mathcal{X}^2_0\mathcal{X}^1_0\rangle\vert\mathcal{A}_0\rangle^2\vert\mathcal{A}_0\rangle^1\vert\mathcal{S}\rangle+ \vert\mathcal{X}^2_0\mathcal{X}^1_1\rangle\vert\mathcal{A}_0\rangle^2\vert\mathcal{A}_0\rangle^1\vert\mathcal{S}\rangle+\vert\mathcal{X}^2_1\mathcal{X}^1_0\rangle\vert\mathcal{A}_0\rangle^2\vert\mathcal{A}_0\rangle^1\vert\mathcal{S}\rangle+\vert\mathcal{X}^2_1\mathcal{X}^1_1\rangle\vert\mathcal{A}_0\rangle^2\vert\mathcal{A}_0\rangle^1\vert\mathcal{S}\rangle\right )
\\= \frac{1}{\sqrt{2^2}} \left (\vert\mathcal{X}^2_0\mathcal{X}^1_0\rangle\vert\mathcal{G}\rangle^{2,1}\vert\mathcal{G}\rangle^{1,1}\vert\mathcal{S}\rangle^{2,1}+\vert\mathcal{X}^2_0\mathcal{X}^1_1\rangle\vert\mathcal{G}\rangle^{2,2}\vert\mathcal{G}\rangle^{1,2}\vert\mathcal{S}\rangle^{2,2}+\vert\mathcal{X}^2_1\mathcal{X}^1_0\rangle\vert\mathcal{G}\rangle^{1,3}\vert\mathcal{G}\rangle^{1,3}\vert\mathcal{S}\rangle^{2,3}+\vert\mathcal{X}^2_1\mathcal{X}^1_1\rangle\vert\mathcal{G}\rangle^{2,4}\vert\mathcal{G}\rangle^{1,4}\vert\mathcal{S}\rangle^{2,4} \right)
    \label{eq:init26}
\end{multline}
Similarly to eq.~\ref{eq:evolutions2}, here let $\vert \mathcal{A}_0\rangle^2\vert\mathcal{A}_0\rangle^1 = \vert 00\rangle$ and $\vert \mathcal{S}\rangle=\vert \mathcal{S}_0\rangle$ . Then $\vert\mathcal{G}\rangle^{2,1}\vert\mathcal{G}\rangle^{1,1}=\vert00\rangle$, $\vert\mathcal{G}\rangle^{2,2}\vert\mathcal{G}\rangle^{1,2}=\vert00\rangle$, $\vert\mathcal{G}\rangle^{2,3}\vert\mathcal{G}\rangle^{1,3}=\vert00\rangle$,  $\vert\mathcal{G}\rangle^{2,4}\vert\mathcal{G}\rangle^{1,4}=\vert00\rangle$ and $\vert\mathcal{S}\rangle^{2,1}=\vert\mathcal{S}_2\rangle$, $\vert\mathcal{S}\rangle^{2,2}=\vert\mathcal{S}_1\rangle$, $\vert\mathcal{S}\rangle^{2,3}=\vert\mathcal{S}_2\rangle$ and $\vert\mathcal{S}\rangle^{2,4}=\vert\mathcal{S}_0\rangle$. The corresponding quantum circuit to eq.~\ref{eq:init26} is shown in Figure~\ref{fig:eQFA1}.

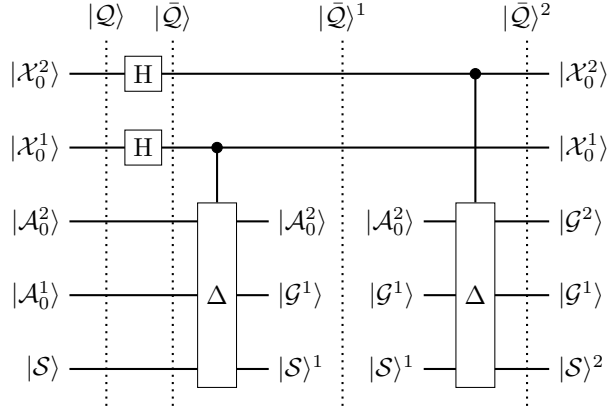
\begin{figure}[bht]
    \centering
    \resizebox{0.5\textwidth}{!}{
\begin{tikzpicture}
\draw[black,thick,dotted]  (-0.5,0.5)node[anchor=north,inner xsep=3pt,inner ysep=1pt,yshift=5.5cm]{$\vert\mathcal{Q}\rangle$} -- (-0.5,5.5);

\draw[black,thick,dotted]  (0.4,0.5)node[anchor=north,inner xsep=3pt,inner ysep=1pt,yshift=5.5cm]{$\vert\bar{\mathcal{Q}}\rangle$} -- (0.4,5.5);

\draw[black,thick,dotted]  (2.7,0.5)node[anchor=north,inner xsep=3pt,inner ysep=1pt,yshift=5.5cm]{$\vert\bar{\mathcal{Q}}\rangle^1$} -- (2.7,5.5);

\draw[black,thick,dotted]  (5.2,0.5)node[anchor=north,inner xsep=3pt,inner ysep=1pt,yshift=5.5cm]{$\vert\bar{\mathcal{Q}}\rangle^2$} -- (5.2,5.5);

\draw[black,thick]  (-1,5)node[anchor=east,inner xsep=3pt,inner ysep=1pt]{$\vert\mathcal{X}^2_0\rangle$} -- (5.5,5)node[anchor=west,inner xsep=3pt,inner ysep=1pt]{$\vert\mathcal{X}^2_0\rangle$};
\draw[black,thick]  (-1,4)node[anchor=east,inner xsep=3pt,inner ysep=1pt]{$\vert \mathcal{X}^1_0\rangle$} -- (5.5,4)node[anchor=west,inner xsep=3pt,inner ysep=1pt]{$\vert\mathcal{X}^1_0\rangle$};

\draw[black,thick]  (-1,3)node[anchor=east,inner xsep=3pt,inner ysep=1pt]{$\vert \mathcal{A}^2_0\rangle$} -- (1.7,3)node[anchor=west,inner xsep=3pt,inner ysep=1pt]{$\vert \mathcal{A}^2_0\rangle$};
\draw[black,thick]  (-1,2)node[anchor=east,inner xsep=3pt,inner ysep=1pt]{$\vert \mathcal{A}^1_0\rangle$}  -- (1.7,2)node[anchor=west,inner xsep=3pt,inner ysep=1pt]{$\vert \mathcal{G}^1\rangle$};
\draw[black,thick]  (-1,1)node[anchor=east,inner xsep=3pt,inner ysep=1pt]{$\vert \mathcal{S}\rangle$} --(1.7,1)node[anchor=west,inner xsep=3pt,inner ysep=1pt]{$\vert \mathcal{S}\rangle^1$};

\draw[black,thick]  (3.8,3)node[anchor=east,inner xsep=3pt,inner ysep=1pt]{$\vert \mathcal{A}^2_0\rangle$} -- (5.5,3)node[anchor=west,inner xsep=3pt,inner ysep=1pt]{$\vert \mathcal{G}^2\rangle$};
\draw[black,thick]  (3.8,2)node[anchor=east,inner xsep=3pt,inner ysep=1pt]{$\vert \mathcal{G}^1\rangle$} -- (5.5,2)node[anchor=west,inner xsep=3pt,inner ysep=1pt]{$\vert \mathcal{G}^1\rangle$};
\draw[black,thick]  (3.8,1)node[anchor=east,inner xsep=3pt,inner ysep=1pt]{$\vert \mathcal{S}\rangle^1$} -- (5.5,1)node[anchor=west,inner xsep=3pt,inner ysep=1pt]{$\vert \mathcal{S}\rangle^2$};

\node[rectangle,draw,fill=white] (h1) at (0,5){H};
\node[rectangle,draw,fill=white] (h2) at (0,4){H};
\node[rectangle,draw,minimum width = 0.5cm, minimum height = 2.5cm,fill=white] (d1) at (1,2) {$\Delta$};
\node[rectangle,draw,minimum width = 0.5cm, minimum height = 2.5cm,fill=white] (d2) at (4.5,2) {$\Delta$};
\filldraw[black] (4.5,5) circle (2pt);
\filldraw[black] (1,4) circle (2pt);

\draw[black,thick]  (1,4)node[anchor=east,inner xsep=3pt,inner ysep=1pt]{} -- (d1.north);
\draw[black,thick]  (4.5,5)node[anchor=east,inner xsep=3pt,inner ysep=1pt]{} -- (d2.north);
\end{tikzpicture}}

    \caption{Circuit corresponding to the FA built from the specifications in Table~\ref{tab:M1}, superposed input qubits $\vert\mathcal{X}^2_0,\mathcal{X}^1_0\rangle$, and two ancillae qubits, one for each sequential input}
    \label{fig:eQFA1}
\end{figure}

The obtained QTS has the sub-QTSs that include the quantum states $\vert \mathcal{S}\rangle^{2,1}$,  $\vert \mathcal{S}\rangle^{2,2}$,  $\vert \mathcal{S}\rangle^{2,3}$ and $\vert \mathcal{S}\rangle^{2,4}$. If the initial QTS contains the quantum state  $\vert \mathcal{S}\rangle = \vert\mathcal{S}_0\rangle$, then  $\vert \mathcal{S}\rangle^1 = \frac{1}{\sqrt{2}}\left (\vert\mathcal{S}\rangle^{1,1}+\vert\mathcal{S}\rangle^{1,2}\right )= \frac{1}{\sqrt{2}}\left (\vert\mathcal{S}_1\rangle+\vert\mathcal{S}_0\rangle\right )$ and $\vert \mathcal{S}\rangle^2=\frac{1}{{2}}\left (\vert\mathcal{S}\rangle^{2,1}+\vert\mathcal{S}\rangle^{2,2}+\vert\mathcal{S}\rangle^{2,3}+\vert\mathcal{S}\rangle^{2,4}\right )=\frac{1}{{2}}\left (\vert\mathcal{S}_2\rangle+\vert\mathcal{S}_1\rangle+\vert\mathcal{S}_2\rangle+\vert\mathcal{S}_0\rangle\right ) = \frac{\sqrt{2}}{{2}}\vert\mathcal{S}_2\rangle+ \frac{1}{{2}}\vert\mathcal{S}_1\rangle+ \frac{1}{{2}}\vert\mathcal{S}_0\rangle$.

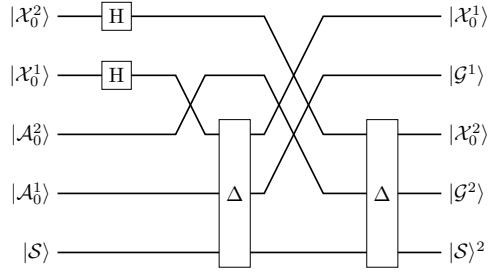
\begin{figure}[bht]
    \centering
    \resizebox{0.4\textwidth}{!}{
\begin{tikzpicture}

\draw[black,thick]  (-1,5)node[anchor=east,inner xsep=3pt,inner ysep=1pt]{$\vert\mathcal{X}^2_0\rangle$} -- (2.5,5) -- (3.5,3) -- (5.5,3)node[anchor=west,inner xsep=3pt,inner ysep=1pt]{$\vert\mathcal{X}^2_0\rangle$};

\draw[black,thick]  (-1,4)node[anchor=east,inner xsep=3pt,inner ysep=1pt]{$\vert \mathcal{X}^1_0\rangle$} -- (1,4) -- (1.5,3) -- (2.5,3) -- (3.5,5) -- (5.5,5)node[anchor=west,inner xsep=3pt,inner ysep=1pt]{$\vert\mathcal{X}^1_0\rangle$}; 

\draw[black,thick]  (-1,3)node[anchor=east,inner xsep=3pt,inner ysep=1pt]{$\vert \mathcal{A}^2_0\rangle$} -- (1,3) -- (1.5,4) -- (2.5,4) -- (3.5,2) -- (5.5,2)node[anchor=west,inner xsep=3pt,inner ysep=1pt]{$\vert \mathcal{G}^2\rangle$};

\draw[black,thick]  (-1,2)node[anchor=east,inner xsep=3pt,inner ysep=1pt]{$\vert \mathcal{A}^1_0\rangle$}  -- (2.5,2) -- (3.5,4) -- (5.5,4)node[anchor=west,inner xsep=3pt,inner ysep=1pt]{$\vert \mathcal{G}^1\rangle$};

\draw[black,thick]  (-1,1)node[anchor=east,inner xsep=3pt,inner ysep=1pt]{$\vert \mathcal{S}\rangle$} -- (5.5,1)node[anchor=west,inner xsep=3pt,inner ysep=1pt]{$\vert \mathcal{S}\rangle^2$};

\node[rectangle,draw,fill=white] (h1) at (0,5){H};
\node[rectangle,draw,fill=white] (h2) at (0,4){H};

\node[rectangle,draw,minimum width = 0.5cm, minimum height = 2.5cm,fill=white] (d1) at (2,2) {$\Delta$};

\node[rectangle,draw,minimum width = 0.5cm, minimum height = 2.5cm,fill=white] (d2) at (4.5,2) {$\Delta$};

\end{tikzpicture}}

    \caption{Circuit showing how the qubits are routed to each of the $\Delta$ operator when used as in circuit in Figure~\ref{fig:eQFA1}}
    \label{fig:deltac}
\end{figure}

The $\Delta$ operator takes only three inputs: the input, the ancilla and the state. Therefore the $\Delta$ operators used when a input sequence are compound operators as shown in circuit in Figure~\ref{fig:deltac}.

\color[rgb]{0,0,0}

\subsection{Executing all possible input sequences starting from all quantum states}

In eq.~\ref{eq:ginitq3} we show how to execute all possible inputs of certain length $l$ starting from one state of the FA and here we generalize the approach considering the simultaneous application of all input sequences starting from each state of the FA. To this end, we map the states of the FA to the corresponding quantum states $\vert\mathcal{S}_n\ldots\mathcal{S}_1\rangle$.


To simultaneously evolve the $n$ FA states, we need $n$ sets of $l$ ancillae indexed by a subscript from $0$ to $n-1$, $\vert\mathcal{A}_0\rangle^{l}_{n-1}\ldots\vert\mathcal{A}_0\rangle^{1}_{n-1}\ldots\vert\mathcal{A}_0\rangle^{l}_0\ldots\vert\mathcal{A}_0\rangle^{1}_0$.
Putting together the superposed input sequences $\vert\prescript{l}{k}{\mathcal{X}}\rangle$ encoded on $l\kappa$ qubits , the ancillae $\vert\mathcal{A}_0\rangle^{l}_{n-1}\ldots\vert\mathcal{A}_0\rangle^{1}_{n-1}\ldots\vert\mathcal{A}_0\rangle^{l}_0\ldots\vert\mathcal{A}_0\rangle^{1}_0$ encoded on $l\nu\mu$ qubits, and the states $\vert\mathcal{S}_{n-1}\ldots\mathcal{S}_0\rangle$ encoded on $n\nu$ qubits, we create the QTS, which contains all the $n$ states of the FA, $\vert\bar{\mathcal{Q}_n}\rangle =\vert\prescript{l}{k}{\mathcal{X}}\rangle\vert\mathcal{A}_0\rangle^{l}_{n-1}\ldots\vert\mathcal{A}_0\rangle^{1}_{n-1}\ldots\vert\mathcal{A}_0\rangle^{l}_0\ldots\vert\mathcal{A}_0\rangle^{1}_0\vert\mathcal{S}_n\ldots\mathcal{S}_1\rangle $.

Then we apply, in parallel, $l$ times in a series the $\Delta$ operator representing the execution of the first, second, to the $l$-th input of each input sequence starting from the initial states of $\vert\bar{\mathcal{Q}_n}\rangle$ as shown in eq.~\ref{eq:result3}.
Because the result is a superposition of all input sequences and all reached FA states, the garbage qubits have been indexed as $\vert\mathcal{G}\rangle_{p}^{l,k^{r}-1}\ldots\vert\mathcal{G}\rangle_{p}^{1,k^{r}-1}$, $r=1..l; p=0...n-1; \sigma=0..k^r-1$. This indexing reflects the fact that the garbage qubits have been obtained by the application of the $\sigma$ input sequence starting from $\vert\mathcal{S}_{p}\rangle$ FA state. Similarly, the reached FA states $\vert\mathcal{S}_{n-1}\ldots\mathcal{S}_0\rangle^{l,k^r-1}$ are indexed so as to reflect that they have been reached by $\sigma$ sequence of length $l$ from the initial states $\vert\mathcal{S}_{n-1}\ldots\mathcal{S}_0\rangle$.

\begin{multline}
    \Delta^l\vert\bar{\mathcal{Q}_n}\rangle =\vert\bar{\mathcal{Q}_n}\rangle^l =\Delta^l\vert\prescript{l}{k}{\mathcal{X}}\rangle\vert\mathcal{A}_0\rangle^{l}_{n-1}\ldots\vert\mathcal{A}_0\rangle^{1}_{n-1}\vert\mathcal{S}_{n-1}\rangle\vert\mathcal{A}_0\rangle^{l}_{n-2}\ldots\vert\mathcal{A}_0\rangle^{1}_{n-2}\vert\mathcal{S}_{n-2}\rangle\ldots\vert\mathcal{A}_0\rangle^{l}_{0}\ldots\vert\mathcal{A}_0\rangle^{1}_{0}\vert\mathcal{S}_0\rangle 
         \\=\frac{1}{\sqrt{k^l}}\left (\vert\mathcal{X}^l_0\ldots\mathcal{X}^1_0\rangle\vert\mathcal{G}\rangle^{l,0}_{n-1}\ldots\vert\mathcal{G}\rangle_{n-1}^{1,0}\vert\mathcal{S}_{n-1}\rangle^{l,0}\dots \vert\mathcal{G}\rangle_0^{l,0}\ldots\vert\mathcal{G}\rangle_{0}^{1,0}\vert\mathcal{S}_{0}\rangle^{l,0}  \right .\\\left .\ldots
         +\vert\mathcal{X}^l_{k^l-1}\ldots\mathcal{X}^1_{k^l-1}\rangle \vert\mathcal{G}\rangle_{n-1}^{l,k^l-1}\ldots\vert\mathcal{G}\rangle_{n-1}^{1,k^l-1}\vert\mathcal{S}_{n-1}\rangle^{l,k^l-1}\dots \vert\mathcal{G}\rangle_0^{l,k^l-1}\ldots\vert\mathcal{G}\rangle_{0}^{1,k^l-1}\vert\mathcal{S}_{0}\rangle^{l,k^l-1}\right )
    \label{eq:result3}
\end{multline}

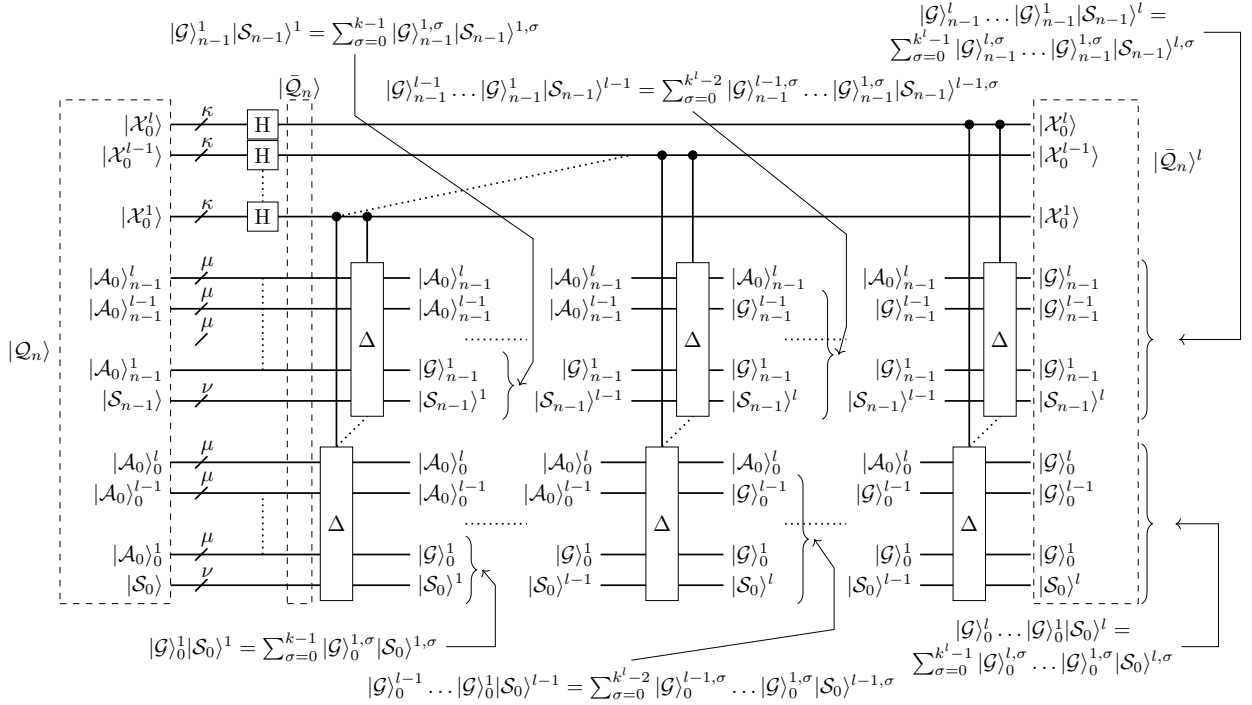
\begin{figure*}[bht]
    \centering
        \resizebox{\textwidth}{!}{

\begin{tikzpicture}

\draw[black,thick](-1.6,8.4) -- (-1.4,8.6)node[anchor=south,yshift=-0.1cm]{$\kappa$};
\draw[black,thick](-1.6,7.9) -- (-1.4,8.1)node[anchor=south,yshift=-0.1cm]{$\kappa$};
\draw[black,thick](-1.6,6.9) -- (-1.4,7.1)node[anchor=south,yshift=-0.1cm]{$\kappa$};
\draw[black,thick](-1.6,5.9) -- (-1.4,6.1)node[anchor=south,yshift=-0.1cm]{$\mu$};
\draw[black,thick](-1.6,5.4) -- (-1.4,5.6)node[anchor=south,yshift=-0.1cm]{$\mu$};
\draw[black,thick](-1.6,4.9) -- (-1.4,5.1)node[anchor=south,yshift=-0.1cm]{$\mu$};
\draw[black,thick](-1.6,3.9) -- (-1.4,4.1)node[anchor=south,yshift=-0.1cm]{$\nu$};
\draw[black,thick](-1.6,2.9) -- (-1.4,3.1)node[anchor=south,yshift=-0.1cm]{$\mu$};
\draw[black,thick](-1.6,2.4) -- (-1.4,2.6)node[anchor=south,yshift=-0.1cm]{$\mu$};
\draw[black,thick](-1.6,1.4) -- (-1.4,1.6)node[anchor=south,yshift=-0.1cm]{$\mu$};
\draw[black,thick](-1.6,0.9) -- (-1.4,1.1)node[anchor=south,yshift=-0.1cm]{$\nu$};

\draw[black,thick]  (-2,6)node[anchor=east,inner xsep=3pt,inner ysep=1pt]{$\vert \mathcal{A}_0\rangle^l_{n-1}$} -- 
(1.9,6)node[anchor=west,inner xsep=3pt,inner ysep=1pt]{$\vert \mathcal{A}_0\rangle^l_{n-1}$};
\draw[black,thick]  (-2,5.5)node[anchor=east,inner xsep=3pt,inner ysep=1pt]{$\vert \mathcal{A}_0\rangle^{l-1}_{n-1}$} -- 
(1.9,5.5)node[anchor=west,inner xsep=3pt,inner ysep=1pt]{$\vert \mathcal{A}_0\rangle^{l-1}_{n-1}$};
\draw[black,thick]  (-2,4.5)node[anchor=east,inner xsep=3pt,inner ysep=1pt]{$\vert \mathcal{A}_0\rangle^1_{n-1}$} -- (1.9,4.5)node[anchor=west,inner xsep=3pt,inner ysep=1pt]{$\vert \mathcal{G}\rangle^1_{n-1}$};
\draw[black,thick]  (-2,3)node[anchor=east,inner xsep=3pt,inner ysep=1pt]{$\vert \mathcal{A}_0\rangle^l_{0}$} -- (1.9,3)node[anchor=west,inner xsep=3pt,inner ysep=1pt]{$\vert \mathcal{A}_0\rangle^l_{0}$};
\draw[black,thick]  (-2,2.5)node[anchor=east,inner xsep=3pt,inner ysep=1pt]{$\vert \mathcal{A}_0\rangle^{l-1}_{0}$} -- (1.9,2.5)node[anchor=west,inner xsep=3pt,inner ysep=1pt]{$\vert \mathcal{A}_0\rangle^{l-1}_{0}$};
\draw[black,thick]  (-2,1.5)node[anchor=east,inner xsep=3pt,inner ysep=1pt]{$\vert \mathcal{A}_0\rangle^1_{0}$} -- (1.9,1.5)node[anchor=west,inner xsep=3pt,inner ysep=1pt]{$\vert \mathcal{G}\rangle^1_{0}$};
\draw[black,thick]  (-2,4)node[anchor=east,inner xsep=3pt,inner ysep=1pt]{$\vert \mathcal{S}_{n-1}\rangle$} -- (1.9,4)node[anchor=west,inner xsep=3pt,inner ysep=1pt]{$\vert \mathcal{S}_{n-1}\rangle^1$};
\draw[black,thick]  (-2,1)node[anchor=east,inner xsep=3pt,inner ysep=1pt]{$\vert \mathcal{S}_0\rangle$} --  (1.9,1)node[anchor=west,inner xsep=3pt,inner ysep=1pt]{$\vert \mathcal{S}_0\rangle^1$};

\draw[black,thick]  (5.5,6)node[anchor=east,inner xsep=3pt,inner ysep=1pt]{$\vert \mathcal{A}_0\rangle^l_{n-1}$} -- (7,6)node[anchor=west,inner xsep=3pt,inner ysep=1pt]{$\vert \mathcal{A}_0\rangle^l_{n-1}$};
\draw[black,thick]  (5.5,5.5)node[anchor=east,inner xsep=3pt,inner ysep=1pt]{$\vert \mathcal{A}_0\rangle^{l-1}_{n-1}$} -- (7,5.5)node[anchor=west,inner xsep=3pt,inner ysep=1pt]{$\vert \mathcal{G}\rangle^{l-1}_{n-1}$};
\draw[black,thick]  (5.5,4.5)node[anchor=east,inner xsep=3pt,inner ysep=1pt]{$\vert \mathcal{G}\rangle^1_{n-1}$} -- (7,4.5)node[anchor=west,inner xsep=3pt,inner ysep=1pt]{$\vert \mathcal{G}\rangle^1_{n-1}$};
\draw[black,thick]  (5.5,4)node[anchor=east,inner xsep=3pt,inner ysep=1pt]{$\vert \mathcal{S}_{n-1}\rangle^{l-1}$} -- (7,4)node[anchor=west,inner xsep=3pt,inner ysep=1pt]{$\vert \mathcal{S}_{n-1}\rangle^l$} ;
\draw[black,thick]  (5,3)node[anchor=east,inner xsep=3pt,inner ysep=1pt]{$\vert \mathcal{A}_0\rangle^l_{0}$} -- (7,3)node[anchor=west,inner xsep=3pt,inner ysep=1pt] (g0l){$\vert \mathcal{A}_0\rangle^l_{0}$};
\draw[black,thick]  (5,2.5)node[anchor=east,inner xsep=3pt,inner ysep=1pt]{$\vert \mathcal{A}_0\rangle^{l-1}_{0}$} -- (7,2.5)node[anchor=west,inner xsep=3pt,inner ysep=1pt] (g0l){$\vert \mathcal{G}\rangle^{l-1}_{0}$};
\draw[black,thick]  (5,1.5)node[anchor=east,inner xsep=3pt,inner ysep=1pt]{$\vert \mathcal{G}\rangle^1_{0}$} -- (7,1.5)node[anchor=west,inner xsep=3pt,inner ysep=1pt] (g00) {$\vert \mathcal{G}\rangle^1_{0}$};
\draw[black,thick]  (5,1)node[anchor=east,inner xsep=3pt,inner ysep=1pt]{$\vert \mathcal{S}_0\rangle^{l-1}$} -- (7,1)node[anchor=west,inner xsep=3pt,inner ysep=1pt]{$\vert \mathcal{S}_0\rangle^l$};

\draw[black,thick]  (10.6,6)node[anchor=east,inner xsep=3pt,inner ysep=1pt]{$\vert \mathcal{A}_0\rangle^l_{n-1}$} -- (12,6)node[anchor=west,inner xsep=3pt,inner ysep=1pt]{$\vert \mathcal{G}\rangle^l_{n-1}$};
\draw[black,thick]  (10.6,5.5)node[anchor=east,inner xsep=3pt,inner ysep=1pt]{$\vert \mathcal{G}\rangle^{l-1}_{n-1}$} -- (12,5.5)node[anchor=west,inner xsep=3pt,inner ysep=1pt]{$\vert \mathcal{G}\rangle^{l-1}_{n-1}$};
\draw[black,thick]  (10.6,4.5)node[anchor=east,inner xsep=3pt,inner ysep=1pt]{$\vert \mathcal{G}\rangle^1_{n-1}$} -- (12,4.5)node[anchor=west,inner xsep=3pt,inner ysep=1pt]{$\vert \mathcal{G}\rangle^1_{n-1}$};
\draw[black,thick]  (10.6,4)node[anchor=east,inner xsep=3pt,inner ysep=1pt]{$\vert \mathcal{S}_{n-1}\rangle^{l-1}$} -- (12,4)node[anchor=west,inner xsep=3pt,inner ysep=1pt]{$\vert \mathcal{S}_{n-1}\rangle^l$} ;
\draw[black,thick]  (10.2,3)node[anchor=east,inner xsep=3pt,inner ysep=1pt]{$\vert \mathcal{A}_0\rangle^l_{0}$} -- (12,3)node[anchor=west,inner xsep=3pt,inner ysep=1pt] (g0l){$\vert \mathcal{G}\rangle^l_{0}$};
\draw[black,thick]  (10.2,2.5)node[anchor=east,inner xsep=3pt,inner ysep=1pt]{$\vert \mathcal{G}\rangle^{l-1}_{0}$} -- (12,2.5)node[anchor=west,inner xsep=3pt,inner ysep=1pt] (g0l){$\vert \mathcal{G}\rangle^{l-1}_{0}$};
\draw[black,thick]  (10.2,1.5)node[anchor=east,inner xsep=3pt,inner ysep=1pt]{$\vert \mathcal{G}\rangle^1_{0}$} -- (12,1.5)node[anchor=west,inner xsep=3pt,inner ysep=1pt] (g00) {$\vert \mathcal{G}\rangle^1_{0}$};
\draw[black,thick]  (10.2,1)node[anchor=east,inner xsep=3pt,inner ysep=1pt]{$\vert \mathcal{S}_0\rangle^{l-1}$} -- (12,1)node[anchor=west,inner xsep=3pt,inner ysep=1pt]{$\vert \mathcal{S}_0\rangle^l$};

\draw[black,thick]  (-2,8.5)node[anchor=east,inner xsep=3pt,inner ysep=1pt]{$\vert \mathcal{X}_0^l\rangle$} -- (12,8.5)node[anchor=west,inner xsep=3pt,inner ysep=1pt]{$\vert \mathcal{X}_0^l\rangle$} ; 
\draw[black,thick]  (-2,8)node[anchor=east,inner xsep=3pt,inner ysep=1pt]{$\vert \mathcal{X}_0^{l-1}\rangle$} -- (12,8)node[anchor=west,inner xsep=3pt,inner ysep=1pt]{$\vert \mathcal{X}_0^{l-1}\rangle$} ; 
\draw[black,thick]  (-2,7)node[anchor=east,inner xsep=3pt,inner ysep=1pt]{$\vert \mathcal{X}_0^1\rangle$} -- (12,7)node[anchor=west,inner xsep=3pt,inner ysep=1pt]{$\vert \mathcal{X}_0^1\rangle$} ;
\node[rectangle,draw,fill=white] (h1) at (-0.5,7){H};
\node[rectangle,draw,fill=white] (h2) at (-0.5,8){H};
\node[rectangle,draw,fill=white] (h3) at (-0.5,8.5){H};

\node[rectangle,draw,minimum width = 0.5cm, minimum height = 2.5cm,fill=white] (d1) at (0.7,2) {$\Delta$};
\node[rectangle,draw,minimum width = 0.5cm, minimum height = 2.5cm,fill=white] (d11) at (1.2,5) {$\Delta$};
\node[rectangle,draw,minimum width = 0.5cm, minimum height = 2.5cm,fill=white] (d12) at (6,2) {$\Delta$};
\node[rectangle,draw,minimum width = 0.5cm, minimum height = 2.5cm,fill=white] (d13) at (6.5,5) {$\Delta$};
\node[rectangle,draw,minimum width = 0.5cm, minimum height = 2.5cm,fill=white] (d2) at (11,2) {$\Delta$};
\node[rectangle,draw,minimum width = 0.5cm, minimum height = 2.5cm,fill=white] (d22) at (11.5,5) {$\Delta$};
\filldraw[black] (11,8.5) circle (2pt);
\filldraw[black] (11.5,8.5) circle (2pt);
\filldraw[black] (6,8) circle (2pt);
\filldraw[black] (6.5,8) circle (2pt);
\filldraw[black] (0.7,7) circle (2pt);
\filldraw[black] (1.2,7) circle (2pt);
\draw[black,thick]  (0.7,7) -- (d1.north);
\draw[black,thick]  (1.2,7) -- (d11.north);
\draw[black,thick]  (6,8) -- (d12.north);
\draw[black,thick]  (6.5,8) -- (d13.north);
\draw[black,thick]  (11,8.5) -- (d2.north);
\draw[black,thick]  (11.5,8.5) -- (d22.north);

\draw[dotted,black,thick](h2.south) -- (h1.north);
\draw[dotted,black,thick](d1.north) -- (d11.south);
\draw[dotted,black,thick](d12.north) -- (d13.south);
\draw[dotted,black,thick](d2.north) -- (d22.south);
\draw[dotted,black,thick](-0.5,4.5) -- (-0.5,6);
\draw[dotted,black,thick](-0.5,1.5) -- (-0.5,2.5);

\draw[dotted,black,thick](0.7,7) -- (5.5,8);
\draw[dotted,black,thick](2.8,2) -- (3.8,2);
\draw[dotted,thick](2.8,5) -- (3.8,5);
\draw[dotted,black,thick](8,2) -- (9,2);
\draw[dotted,thick](8,5) -- (9,5);

\node[rectangle,draw,minimum width = 1.8cm, minimum height = 8.2cm,dashed] (r1l) at (-2.9,4.8){};
\node[left of=r1l,xshift=-0.4cm] (r1ll) {$\vert\mathcal{Q}_n\rangle$};
\node[rectangle,draw,minimum width = 0.4cm, minimum height = 8.2cm,dashed] (rl) at (0.1,4.8){};
\node[above of=rl,yshift=3.3cm] (rll) {$\vert\bar{\mathcal{Q}}_n\rangle$};
\node[rectangle,draw,minimum width = 1.7cm, minimum height = 8.2cm,dashed] (rfl) at (12.9,4.8){};
\node[above of=rfl,yshift=2.1cm,xshift=1.5cm] (rfll) {$\vert\bar{\mathcal{Q}}_n\rangle^l$};

\draw [decorate,decoration={brace,amplitude=5pt,mirror,raise=4ex}] (2.2,0.7) -- (2.2,1.8){} node[midway,xshift=1cm] (fl1) {};
\node[] (fll1) at (0,0)  {$\vert\mathcal{G}\rangle^1_{0}\vert\mathcal{S}_0\rangle^{1} = \sum _{\sigma=0}^{k-1}\vert\mathcal{G}\rangle_{0}^{1,\sigma}\vert\mathcal{S}_0\rangle^{1,\sigma}$};
\draw[black, ->] (fll1.east) -- ++(0.8,0) -- ++(0,1) -- (fl1.west);

\draw [decorate,decoration={brace,amplitude=5pt,mirror,raise=4ex}] (2.8,3.7) -- (2.8,4.8){} node[midway,xshift=1cm] (ffl1) {};
\node[] (ffll1) at (1,10)  {$\vert\mathcal{G}\rangle^1_{n-1}\vert\mathcal{S}_{n-1}\rangle^{1} = \sum _{\sigma=0}^{k-1}\vert\mathcal{G}\rangle_{n-1}^{1,\sigma}\vert\mathcal{S}_{n-1}\rangle^{1,\sigma}$};
\draw[black, ->] (ffll1.south) -- ++(0,-1) -- ++(2.9,-2) -- ++(0,-2) -- (ffl1.west);

\draw [decorate,decoration={brace,amplitude=5pt,mirror,raise=4ex}] (7.6,0.7) -- (7.6,2.8){} node[midway,xshift=1cm] (fl2) {};
\node[] (fll2) at (5.5,-0.6)  {$\vert\mathcal{G}\rangle_{0}^{l-1}\ldots\vert\mathcal{G}\rangle_{0}^1\vert\mathcal{S}_{0}\rangle^{l-1} = \sum _{\sigma=0}^{k^l-2}\vert\mathcal{G}\rangle_{0}^{l-1,\sigma}\ldots\vert\mathcal{G}\rangle_{0}^{1,\sigma}\vert\mathcal{S}_0\rangle^{l-1,\sigma}$};
\draw[black, ->] (fll2.north) -- ++(3.3,0.5) -- ++(0,0.2)  -- ++(0,0.8) -- (fl2.west);

\draw [decorate,decoration={brace,amplitude=5pt,mirror,raise=4ex}] (8,3.7) -- (8,5.8){} node[midway,xshift=1cm] (ffl2) {};
\node[] (ffll2) at (6.5,9.1)  {$\vert\mathcal{G}\rangle_{n-1}^{l-1}\ldots\vert\mathcal{G}\rangle_{n-1}^1\vert\mathcal{S}_{n-1}\rangle^{l-1} =\sum _{\sigma=0}^{k^l-2}\vert\mathcal{G}\rangle_{n-1}^{l-1,\sigma}\ldots\vert\mathcal{G}\rangle_{n-1}^{1,\sigma}\vert\mathcal{S}_{n-1}\rangle^{l-1,\sigma}$};
\draw[black, ->] (ffll2.south) -- ++(2.5,-2) -- ++(0,-1.2)  -- ++(0,-0.3) -- (ffl2.west);

\draw [decorate,decoration={brace,amplitude=5pt,mirror,raise=4ex}]
  (13.2,3.7) -- (13.2,6.3){} node[midway,yshift=5cm,xshift=-1cm,align=center]  (oo1) {$\vert\mathcal{G}\rangle_{n-1}^{l}\ldots\vert\mathcal{G}\rangle_{n-1}^1\vert\mathcal{S}_{n-1}\rangle^{l} =$\\$\sum _{\sigma=0}^{k^l-1}\vert\mathcal{G}\rangle_{n-1}^{l,\sigma}\ldots\vert\mathcal{G}\rangle_{n-1}^{1,\sigma}\vert\mathcal{S}_{n-1}\rangle^{l,\sigma}$};
\draw[black, ->] (oo1.east) -- ++(0.6,0) -- ++(0,-5)  -- ++(-1,0);

\draw [decorate,decoration={brace,amplitude=5pt,mirror,raise=4ex}]
  (13.2,0.7) -- (13.2,3.3){} node[midway,yshift=-2cm,xshift=-1cm,align=center]  (oo2) {$\vert\mathcal{G}\rangle_{0}^{l}\ldots\vert\mathcal{G}\rangle_{0}^1\vert\mathcal{S}_{0}\rangle^{l} =$\\$\sum _{\sigma=0}^{k^l-1}\vert\mathcal{G}\rangle_{0}^{l,\sigma}\ldots\vert\mathcal{G}\rangle_{0}^{1,\sigma}\vert\mathcal{S}_{0}\rangle^{l,\sigma}$};
\draw[black, ->] (oo2.east) -- ++(0.6,0) -- ++(0,2)  -- ++(-0.7,0);

\end{tikzpicture}}
    \caption{
    A circuit representing the applications of all input sequences of length $l$ in supposition; namely, applying in order the first, second, .., $l-th$ inputs of these sequences to all state of the FA encoded in quantum states $\vert\mathcal{S}_{0}\rangle$,..., $\vert\mathcal{S}_{n-1}\rangle$.
    }
    \label{fig:fsms25}
\end{figure*}

The circuit corresponding to eq.~\ref{eq:result3} 
is shown in Figure~\ref{fig:fsms25}. 
Each quantum state $\vert\mathcal{G}\rangle_p^{r}\ldots\vert\mathcal{G}\rangle^1_p\vert\mathcal{S}_p\rangle^{r} = \sum_{\sigma=0}^{k^r-1}\vert\mathcal{G}\rangle_{p}^{r,\sigma} \ldots\vert\mathcal{G}\rangle_{p}^{1,\sigma}\vert\mathcal{S}_p\rangle^{r,\sigma}$ 
contains the FA states encoded by 
$\vert\mathcal{S}_p\rangle^{r}$, reached by the application of all input sequences of length $r$ to the FA states encoded by $\vert\mathcal{S}_p\rangle$. 

\color[rgb]{0,0,0}
\textit{Working example}: The example FA has two states and thus the initial TQS for such a FA is shown in eq.~\ref{eq:init351}. 
\begin{equation}
    \vert\mathcal{Q}_2\rangle =\vert\mathcal{X}_0^2\mathcal{X}_0^1\rangle\vert \mathcal{A}_0\rangle^2_2\vert\mathcal{A}_0\rangle^1_2\vert \mathcal{A}_0\rangle^2_1\vert\mathcal{A}_0\rangle^1_1\vert\mathcal{A}_0\rangle^2_0\vert\mathcal{A}_0\rangle^1_0\vert\mathcal{S}_2\mathcal{S}_1\mathcal{S}_0\rangle
    \label{eq:init351}
\end{equation}
We index the ancillae by a subscript $q=0,1$ such that $\vert\mathcal{A}^l_0\rangle_q$ represents an ancilla that will be used with the $q^{th}$ initial FA state $\vert\mathcal{S}_q\rangle$.

Following eq.~\ref{eq:exqbar}, we put the inputs in eq.~\ref{eq:init351} in superposition resulting in the quantum state $\vert\bar{\mathcal{Q}}\rangle$ shown in eq.~\ref{eq:init35}.
\begin{multline}
    \vert\bar{\mathcal{Q}}_2\rangle
    =\vert\prescript{2}{2}{\mathcal{X}}\rangle\vert \mathcal{A}_0\rangle^2_2\vert\mathcal{A}_0\rangle^1_2\vert \mathcal{A}_0\rangle^2_1\vert\mathcal{A}_0\rangle^1_1\vert\mathcal{A}_0\rangle^2_0\vert\mathcal{A}_0\rangle^1_0\vert\mathcal{S}_2\mathcal{S}_1\mathcal{S}_0\rangle\\=\vert\prescript{2}{2}{\mathcal{X}}\rangle\vert \mathcal{A}_0\rangle^2_2\vert\mathcal{A}_0\rangle^1_2\vert\mathcal{S}_2\rangle\vert \mathcal{A}_0\rangle^2_1\vert\mathcal{A}_0\rangle^1_1\vert\mathcal{S}_1\rangle\vert\mathcal{A}_0\rangle^2_0\vert\mathcal{A}_0\rangle^1_0\vert\mathcal{S}_0\rangle
    \label{eq:init35}
\end{multline}
%

The two consecutive applications of $\Delta$ (similar to eq.~\ref{eq:init26}) to 
$\vert \bar{Q}\rangle$ from eq.~\ref{eq:init35} are shown in eq.~\ref{eq:init4}.

\begin{multline}
\Delta^{2}\vert\bar{\mathcal{Q}}_2\rangle=\vert\bar{\mathcal{Q}}_2\rangle^2= \Delta^{2}\vert\prescript{2}{2}{\mathcal{X}}\rangle\vert \mathcal{A}_0\rangle^2_2\vert\mathcal{A}_0\rangle^1_2\vert\mathcal{S}_2\rangle \mathcal{A}_0\rangle^2_1\vert\mathcal{A}_0\rangle^1_1\vert\mathcal{S}_1\rangle\vert\mathcal{A}_0\rangle^2_0\vert\mathcal{A}_0\rangle^1_0\vert\mathcal{S}_0\rangle\\= \frac{1}{\sqrt{2^2}}\left [ \Delta\Delta\vert \mathcal{A}_0\rangle^2_2\vert\mathcal{A}_0\rangle^1_2\vert\mathcal{S}_2\rangle\vert  \mathcal{A}_0\rangle^2_1\vert\mathcal{A}_0\rangle^1_1\vert\mathcal{S}_1\rangle\vert\mathcal{A}_0\rangle^2_0\vert\mathcal{A}_0\rangle^1_0\vert\mathcal{S}_0\rangle +\Delta\Delta\vert \mathcal{A}_0\rangle^2_2\vert\mathcal{A}_0\rangle^1_2\vert\mathcal{S}_2\rangle\vert  \mathcal{A}_0\rangle^2_1\vert\mathcal{A}_0\rangle^1_1\vert\mathcal{S}_1\rangle\vert\mathcal{A}_0\rangle^2_0\vert\mathcal{A}_0\rangle^1_0\vert\mathcal{S}_0\rangle \right . \\ \left .+\Delta\Delta\vert \mathcal{A}_0\rangle^2_2\vert\mathcal{A}_0\rangle^1_2\vert\mathcal{S}_2\rangle \vert \mathcal{A}_0\rangle^2_1\vert\mathcal{A}_0\rangle^1_1\vert\mathcal{S}_1\rangle\vert\mathcal{A}_0\rangle^2_0\vert\mathcal{A}_0\rangle^1_0\vert\mathcal{S}_0\rangle \Delta\Delta\vert \mathcal{A}_0\rangle^2_2\vert\mathcal{A}_0\rangle^1_2\vert\mathcal{S}_2\rangle\vert  \mathcal{A}_0\rangle^2_1\vert\mathcal{A}_0\rangle^1_1\vert\mathcal{S}_1\rangle\vert\mathcal{A}_0\rangle^2_0\vert\mathcal{A}_0\rangle^1_0\vert\mathcal{S}_0\rangle\right ]
\\= \frac{1}{\sqrt{2^2}}\left [ \vert \mathcal{X}_0^2\mathcal{X}_0^1\rangle\vert \mathcal{G}\rangle_2^{2,1}\vert\mathcal{G}\rangle_2^{1,1}\vert\mathcal{S}_2\rangle^{2,1}\vert \mathcal{G}\rangle_1^{2,1} \vert\mathcal{G}\rangle_1^{1,1}\vert\mathcal{S}_1\rangle^{2,1}\vert\mathcal{G}\rangle_0^{2,1}\vert\mathcal{G}\rangle_0^{1,1}\vert\mathcal{S}_0\rangle^{2,1} \right . \\ \left .+\vert \mathcal{X}_0^2\mathcal{X}_1^1\rangle\vert \mathcal{G}\rangle_2^{2,2}\vert\mathcal{G}\rangle_2^{1,2}\vert\mathcal{S}_2\rangle^{2,2}\vert \mathcal{G}\rangle_1^{2,2} \vert\mathcal{G}1\rangle_1^{1,2}\vert\mathcal{S}_1\rangle^{2,2}\vert\mathcal{G}\rangle_0^{2,2}\vert\mathcal{G}\rangle_0^{1,2}\vert\mathcal{S}_0\rangle^{2,2} \right . \\ \left .+\vert \mathcal{X}_1^2\mathcal{X}_0^1\rangle\vert \mathcal{G}\rangle_2^{2,3}\vert\mathcal{G}\rangle_2^{1,3}\vert\mathcal{S}_2\rangle^{2,3}\vert \mathcal{G}\rangle_1^{2,3} \vert\mathcal{G}\rangle_1^{1,3}\vert\mathcal{S}_1\rangle^{2,3}\vert\mathcal{G}\rangle_0^{2,3}\vert\mathcal{G}\rangle_0^{1,3}\vert\mathcal{S}_0\rangle^{2,3} \right . \\ \left .+\vert \mathcal{X}_1^2\mathcal{X}_1^1\rangle\vert \mathcal{G}\rangle_2^{2,4}\vert\mathcal{G}\rangle_2^{1,4}\vert\mathcal{S}_2\rangle^{2,4}\vert \mathcal{G}\rangle_1^{2,4} \vert\mathcal{G}\rangle_1^{1,4}\vert\mathcal{S}_1\rangle^{2,4}\vert\mathcal{G}\rangle_0^{2,4}\vert\mathcal{G}\rangle_0^{1,4}\vert\mathcal{S}_0\rangle^{2,4} \right ]
\label{eq:init4}
\end{multline}

with $\vert \mathcal{G}\rangle_2^{2,1}\vert\mathcal{G}\rangle_2^{1,1}\vert\mathcal{S}_2\rangle^{2,1}=\vert 00\rangle\vert\mathcal{S}_1\rangle$, $\vert \mathcal{G}\rangle_1^{2,1}\vert\mathcal{G}\rangle_1^{1,1}\vert\mathcal{S}_1\rangle^{2,1}=\vert 00\rangle\vert\mathcal{S}_0\rangle$, $\vert\mathcal{G}\rangle_0^{2,1}\vert\mathcal{G}\rangle_0^{1,1}\vert\mathcal{S}_0\rangle^{2,1}=\vert00\rangle\vert\mathcal{S}_2\rangle$, $\vert\mathcal{G}\rangle_2^{2,2}\vert\mathcal{G}\rangle_2^{1,2}\vert\mathcal{S}_2\rangle^{2,2}=\vert01\rangle\vert\mathcal{S}_0\rangle$, $\vert\mathcal{G}\rangle_1^{2,2}\vert\mathcal{G}\rangle_1^{1,2}\vert\mathcal{S}_1\rangle^{2,2}=\vert00\rangle\vert\mathcal{S}_0\rangle$, $\vert\mathcal{G}\rangle_0^{2,2}\vert\mathcal{G}\rangle_0^{1,2}\vert\mathcal{S}_0\rangle^{2,2}=\vert00\rangle\vert\mathcal{S}_1\rangle$, $\vert\mathcal{G}\rangle_2^{2,3}\vert\mathcal{G}\rangle_2^{1,3}\vert\mathcal{S}_2\rangle^{2,3}=\vert00\rangle\vert\mathcal{S}_0\rangle$, $\vert\mathcal{G}\rangle_1^{2,3}\vert\mathcal{G}\rangle_1^{1,3}\vert\mathcal{S}_1\rangle^{2,3}=\vert10\rangle\vert\mathcal{S}_2\rangle$, $\vert\mathcal{G}\rangle_0^{2,3}\vert\mathcal{G}\rangle_0^{1,3}\vert\mathcal{S}_0\rangle^{2,3}=\vert00\rangle\vert\mathcal{S}_2\rangle$, $\vert \mathcal{G}\rangle_2^{2,4}\vert\mathcal{G}\rangle_2^{1,4}\vert\mathcal{S}_2\rangle^{2,4}=\vert 11\rangle\vert\mathcal{S}_2\rangle$, $\vert \mathcal{G}\rangle_1^{2,4}\vert\mathcal{G}\rangle_1^{1,4}\vert\mathcal{S}_1\rangle^{2,4}=\vert 10\rangle\vert\mathcal{S}_2\rangle$, $\vert\mathcal{G}\rangle_0^{2,4}\vert\mathcal{G}\rangle_0^{1,4}\vert\mathcal{S}_0\rangle^{2,4}=\vert00\rangle\vert\mathcal{S}_0\rangle$. 
However to the shortest resetting sequence has length $l=4$. To save space we show only their resulting QTS with the input sequences and the state collections while omitting the garbage qubits. 

\begin{multline}
\frac{1}{\sqrt{2^4}}\left ( \vert \mathcal{X}_0^4\mathcal{X}_0^3\mathcal{X}_0^2\mathcal{X}_0^1\rangle\vert\mathcal{S}_0\rangle\vert\mathcal{S}_1\rangle\vert\mathcal{S}_2\rangle+[\vert \mathcal{X}_0^4\mathcal{X}_0^3\mathcal{X}_0^2\mathcal{X}_1^1\rangle+\vert \mathcal{X}_0^4\mathcal{X}_1^3\mathcal{X}_0^2\mathcal{X}_0^1\rangle+\vert \mathcal{X}_0^4\mathcal{X}_1^3\mathcal{X}_1^2\mathcal{X}_0^1\rangle+\vert \mathcal{X}_0^4\mathcal{X}_1^3\mathcal{X}_1^2\mathcal{X}_1^1\rangle]\vert\mathcal{S}_0\rangle\vert\mathcal{S}_0\rangle\vert\mathcal{S}_1\rangle\right . \\ \left .+[\vert \mathcal{X}_0^4\mathcal{X}_0^3\mathcal{X}_1^2\mathcal{X}_0^1\rangle+\vert \mathcal{X}_0^4\mathcal{X}_0^3\mathcal{X}_1^2\mathcal{X}_1^1\rangle]\vert\mathcal{S}_1\rangle\vert\mathcal{S}_1\rangle\vert\mathcal{S}_2\rangle +\vert \mathcal{X}_0^4\mathcal{X}_1^3\mathcal{X}_0^2\mathcal{X}_1^1\rangle\vert\mathcal{S}_0\rangle\vert\mathcal{S}_1\rangle\vert\mathcal{S}_1\rangle \right . \\ \left .+(\vert \mathcal{X}_1^4\mathcal{X}_0^3\mathcal{X}_0^2\mathcal{X}_0^1\rangle+\vert \mathcal{X}_1^4\mathcal{X}_1^3\mathcal{X}_0^2\mathcal{X}_0^1\rangle+\vert \mathcal{X}_1^4\mathcal{X}_1^3\mathcal{X}_1^2\mathcal{X}_0^1\rangle+\vert \mathcal{X}_1^4\mathcal{X}_1^3\mathcal{X}_1^2\mathcal{X}_1^1\rangle)\vert\mathcal{S}_0\rangle\vert\mathcal{S}_2\rangle\vert\mathcal{S}_2\rangle\right . \\ \left .+\vert \mathcal{X}_1^4\mathcal{X}_0^3\mathcal{X}_0^2\mathcal{X}_1^1\rangle\vert\mathcal{S}_2\rangle\vert\mathcal{S}_2\rangle\vert\mathcal{S}_2\rangle +[\vert \mathcal{X}_1^4\mathcal{X}_0^3\mathcal{X}_1^2\mathcal{X}_0^1\rangle+\vert \mathcal{X}_1^4\mathcal{X}_0^3\mathcal{X}_1^2\mathcal{X}_1^1\rangle+\vert \mathcal{X}_1^4\mathcal{X}_1^3\mathcal{X}_0^2\mathcal{X}_1^1\rangle] \vert\mathcal{S}_0\rangle\vert\mathcal{S}_0\rangle\vert\mathcal{S}_2\rangle\right )
\label{eq:init8}
\end{multline}

which concretely results in 

\begin{multline}
\frac{1}{\sqrt{2^4}}\left ( \vert 0000\rangle\vert 012\rangle +(\vert 0001\rangle+\vert 0100\rangle+\vert 0110\rangle+\vert 0111\rangle)\vert 001\rangle +(\vert 0010\rangle+\vert 0011\rangle)\vert 112\rangle+\vert 0101\rangle\vert 011\rangle \right . \\ \left .+(\vert 1000\rangle+\vert 1100\rangle+\vert 1110\rangle+\vert 1111\rangle)\vert 022\rangle+\vert 1001\rangle\vert 222\rangle +(\vert 1010\rangle+\vert 1011\rangle+\vert 1101\rangle)\vert 002\rangle\right )
\label{eq:init85}
\end{multline}

The corresponding circuit is shown in Figure~\ref{fig:fsms25}. 

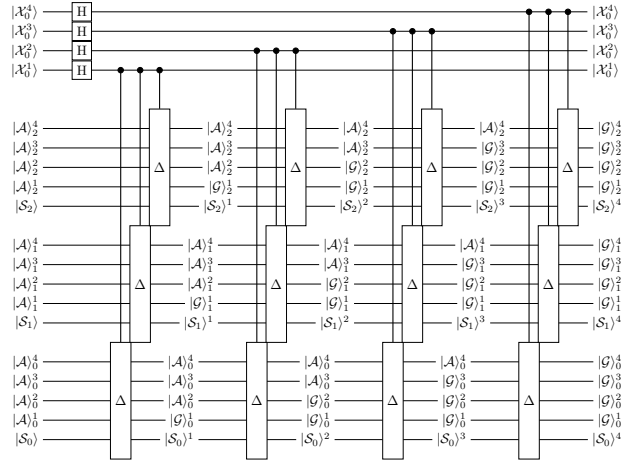
\begin{figure*}[bht]
    \centering
        \resizebox{0.5\textwidth}{!}{
\begin{tikzpicture}

\def\axs{{-1,3,6.5,10,14}};
\def\bxs{{-1,3.5,7,10.5,14}};
\def\cxs{{-1,4,7.5,11,14}};
\def\steps{{0,1,2,3,4}};
\def\ya{-4};
\def\yb{-1};
\def\yc{2};
\def\la{{0,1,2}};
\draw[black] (-1,7)node[anchor=east]{$\vert\mathcal{X}_0^4\rangle$} -- (13,7)node[anchor=west]{$\vert\mathcal{X}_0^4\rangle$};
\draw[black] (-1,6.5)node[anchor=east]{$\vert\mathcal{X}_0^3\rangle$} -- (13,6.5)node[anchor=west]{$\vert\mathcal{X}_0^3\rangle$};
\draw[black] (-1,6)node[anchor=east]{$\vert\mathcal{X}_0^2\rangle$} -- (13,6)node[anchor=west]{$\vert\mathcal{X}_0^2\rangle$};
\draw[black] (-1,5.5)node[anchor=east]{$\vert\mathcal{X}_0^1\rangle$} -- (13,5.5)node[anchor=west]{$\vert\mathcal{X}_0^1\rangle$};

  \foreach \i in {1,...,4} {
    \node[anchor=east,inner xsep=3pt,inner ysep=1pt] (a\i) at (\axs[\i],\ya) {$\vert \mathcal{S}_{\pgfmathparse{\la[0]}\pgfmathresult}\rangle^{\pgfmathparse{\i}\pgfmathresult}$};
    \node[anchor=east,inner xsep=3pt,inner ysep=1pt] (aa\i) at (\axs[\i],\ya+0.5) {$\vert \mathcal{G}\rangle_{\pgfmathparse{\la[0]}\pgfmathresult}^1$};    
    \node[anchor=east,inner xsep=3pt,inner ysep=1pt] (b\i) at (\bxs[\i],\yb) {$\vert \mathcal{S}_{\pgfmathparse{\la[1]}\pgfmathresult}\rangle^{\pgfmathparse{\i}\pgfmathresult}$};    
    \node[anchor=east,inner xsep=3pt,inner ysep=1pt] (bb\i) at (\bxs[\i],\yb+0.5) {$\vert \mathcal{G}\rangle_{\pgfmathparse{\la[1]}\pgfmathresult}^1$};        
    \node[anchor=east,inner xsep=3pt,inner ysep=1pt] (c\i) at (\cxs[\i],\yc) {$\vert \mathcal{S}_{\pgfmathparse{\la[2]}\pgfmathresult}\rangle^{\pgfmathparse{\i}\pgfmathresult}$};    
    \node[anchor=east,inner xsep=3pt,inner ysep=1pt] (cc\i) at (\cxs[\i],\yc+0.5) {$\vert \mathcal{G}\rangle_{\pgfmathparse{\la[2]}\pgfmathresult}^1$};       
  }

  \foreach \i in {2,...,4} {
    \node[anchor=east,inner xsep=3pt,inner ysep=1pt] (aaa\i) at (\axs[\i],\ya+1) {$\vert \mathcal{G}\rangle_{\pgfmathparse{\la[0]}\pgfmathresult}^{2}$};    
    \node[anchor=east,inner xsep=3pt,inner ysep=1pt] (bbb\i) at (\bxs[\i],\yb+1) {$\vert \mathcal{G}\rangle_{\pgfmathparse{\la[1]}\pgfmathresult}^2$};        
    \node[anchor=east,inner xsep=3pt,inner ysep=1pt] (ccc\i) at (\cxs[\i],\yc+1) {$\vert \mathcal{G}\rangle_{\pgfmathparse{\la[2]}\pgfmathresult}^2$};       
  }
  \foreach \i in {3,...,4} {
    \node[anchor=east,inner xsep=3pt,inner ysep=1pt] (aaaa\i) at (\axs[\i],\ya+1.5) {$\vert \mathcal{G}\rangle_{\pgfmathparse{\la[0]}\pgfmathresult}^{3}$};    
    \node[anchor=east,inner xsep=3pt,inner ysep=1pt] (bbbb\i) at (\bxs[\i],\yb+1.5) {$\vert \mathcal{G}\rangle_{\pgfmathparse{\la[1]}\pgfmathresult}^{3}$};        
    \node[anchor=east,inner xsep=3pt,inner ysep=1pt] (cccc\i) at (\cxs[\i],\yc+1.5) {$\vert \mathcal{G}\rangle_{\pgfmathparse{\la[2]}\pgfmathresult}^{3}$};       
  }
  \foreach \i in {4,...,4} {
    \node[anchor=east,inner xsep=3pt,inner ysep=1pt] (aaaaa\i) at (\axs[\i],\ya+2) {$\vert \mathcal{G}\rangle_{\pgfmathparse{\la[0]}\pgfmathresult}^{4}$};    
    \node[anchor=east,inner xsep=3pt,inner ysep=1pt] (bbbbb\i) at (\bxs[\i],\yb+2) {$\vert \mathcal{G}\rangle_{\pgfmathparse{\la[1]}\pgfmathresult}^{4}$};        
    \node[anchor=east,inner xsep=3pt,inner ysep=1pt] (ccccc\i) at (\cxs[\i],\yc+2) {$\vert \mathcal{G}\rangle_{\pgfmathparse{\la[2]}\pgfmathresult}^{4}$};       
  }
  \foreach \i in {0,...,0} {
    \node[anchor=east,inner xsep=3pt,inner ysep=1pt] (a\i) at (\axs[\i],\ya) {$\vert \mathcal{S}_{\pgfmathparse{\la[0]}\pgfmathresult}\rangle$};
    \node[anchor=east,inner xsep=3pt,inner ysep=1pt] (aa\i) at (\axs[\i],\ya+0.5) {$\vert \mathcal{A}\rangle_{\pgfmathparse{\la[0]}\pgfmathresult}^{1}$};    
    \node[anchor=east,inner xsep=3pt,inner ysep=1pt] (b\i) at (\bxs[\i],\yb) {$\vert \mathcal{S}_{\pgfmathparse{\la[1]}\pgfmathresult}\rangle$};    
    \node[anchor=east,inner xsep=3pt,inner ysep=1pt] (bb\i) at (\bxs[\i],\yb+0.5) {$\vert \mathcal{A}\rangle_{\pgfmathparse{\la[1]}\pgfmathresult}^{1}$};        
    \node[anchor=east,inner xsep=3pt,inner ysep=1pt] (c\i) at (\cxs[\i],\yc) {$\vert \mathcal{S}_{\pgfmathparse{\la[2]}\pgfmathresult}\rangle$};    
    \node[anchor=east,inner xsep=3pt,inner ysep=1pt] (cc\i) at (\cxs[\i],\yc+0.5) {$\vert \mathcal{A}\rangle_{\pgfmathparse{\la[2]}\pgfmathresult}^{1}$};       
  }
  \foreach \i in {0,...,1} {
    \node[anchor=east,inner xsep=3pt,inner ysep=1pt] (aaa\i) at (\axs[\i],\ya+1) {$\vert \mathcal{A}\rangle_{\pgfmathparse{\la[0]}\pgfmathresult}^{2}$};    
    \node[anchor=east,inner xsep=3pt,inner ysep=1pt] (bbb\i) at (\bxs[\i],\yb+1) {$\vert \mathcal{A}\rangle_{\pgfmathparse{\la[1]}\pgfmathresult}^{2}$};        
    \node[anchor=east,inner xsep=3pt,inner ysep=1pt] (ccc\i) at (\cxs[\i],\yc+1) {$\vert \mathcal{A}\rangle_{\pgfmathparse{\la[2]}\pgfmathresult}^{2}$};       
  }
  \foreach \i in {0,...,2} {
    \node[anchor=east,inner xsep=3pt,inner ysep=1pt] (aaaa\i) at (\axs[\i],\ya+1.5) {$\vert \mathcal{A}\rangle_{\pgfmathparse{\la[0]}\pgfmathresult}^{3}$};    
    \node[anchor=east,inner xsep=3pt,inner ysep=1pt] (bbbb\i) at (\bxs[\i],\yb+1.5) {$\vert \mathcal{A}\rangle_{\pgfmathparse{\la[1]}\pgfmathresult}^{3}$};        
    \node[anchor=east,inner xsep=3pt,inner ysep=1pt] (cccc\i) at (\cxs[\i],\yc+1.5) {$\vert \mathcal{A}\rangle_{\pgfmathparse{\la[2]}\pgfmathresult}^{3}$};       
  }  
  \foreach \i in {0,...,3} {
    \node[anchor=east,inner xsep=3pt,inner ysep=1pt] (aaaaa\i) at (\axs[\i],\ya+2) {$\vert \mathcal{A}\rangle_{\pgfmathparse{\la[0]}\pgfmathresult}^{4}$};    
    \node[anchor=east,inner xsep=3pt,inner ysep=1pt] (bbbbb\i) at (\bxs[\i],\yb+2) {$\vert \mathcal{A}\rangle_{\pgfmathparse{\la[1]}\pgfmathresult}^{4}$};        
    \node[anchor=east,inner xsep=3pt,inner ysep=1pt] (ccccc\i) at (\cxs[\i],\yc+2) {$\vert \mathcal{A}\rangle_{\pgfmathparse{\la[2]}\pgfmathresult}^{4}$};       
  }  
  \foreach \i in {0,...,3}{
  \pgfmathint{\i+1}\pgfmathsetmacro{\y}{\pgfmathresult};
    \path[draw] (a\i.east) -- (a\y.west);  
    \path[draw] (b\i.east) -- (b\y.west);  
    \path[draw] (c\i.east) -- (c\y.west);
    \path[draw] (aa\i.east) -- (aa\y.west);  
    \path[draw] (bb\i.east) -- (bb\y.west);  
    \path[draw] (cc\i.east) -- (cc\y.west);    
    \path[draw] (aaa\i.east) -- (aaa\y.west);  
    \path[draw] (bbb\i.east) -- (bbb\y.west);  
    \path[draw] (ccc\i.east) -- (ccc\y.west);    
    \path[draw] (aaaa\i.east) -- (aaaa\y.west);  
    \path[draw] (bbbb\i.east) -- (bbbb\y.west);  
    \path[draw] (cccc\i.east) -- (cccc\y.west);        
    \path[draw] (aaaaa\i.east) -- (aaaaa\y.west);  
    \path[draw] (bbbbb\i.east) -- (bbbbb\y.west);  
    \path[draw] (ccccc\i.east) -- (ccccc\y.west);            
    }




\node[rectangle,draw,fill=white] (h1) at (0,5.5){H};
\node[rectangle,draw,fill=white] (h2) at (0,6){H};
\node[rectangle,draw,fill=white] (h3) at (0,6.5){H};
\node[rectangle,draw,fill=white] (h4) at (0,7){H};
\def\ds{{1,4.5,8,11.5}}
\def\ins{{5.5,6,6.5,7}}
  \foreach \i in {0,...,3} {
\node[rectangle,draw,minimum width = 0.5cm, minimum height = 3cm,fill=white] (d\i0) at (\ds[\i],-3) {$\Delta$};
\node[rectangle,draw,minimum width = 0.5cm, minimum height = 3cm,fill=white] (d\i1) at (\ds[\i]+0.5,0) {$\Delta$};
\node[rectangle,draw,minimum width = 0.5cm, minimum height = 3cm,fill=white] (d\i2) at (\ds[\i]+1,3) {$\Delta$};
\filldraw[black] (\ds[\i],\ins[\i]) circle (2pt);
\filldraw[black] (\ds[\i]+0.5,\ins[\i]) circle (2pt);
\filldraw[black] (\ds[\i]+1,\ins[\i]) circle (2pt);
\draw[black,thick]  (\ds[\i],\ins[\i]) -- (d\i0.north);
\draw[black,thick]  (\ds[\i]+0.5,\ins[\i]) -- (d\i1.north);
\draw[black,thick]  (\ds[\i]+1,\ins[\i]) -- (d\i2.north);
}
\end{tikzpicture}}
    \caption{
    A circuit representing the application of all input sequences of length $2$ to all states of the FA. }
    \label{fig:fsms25}
\end{figure*}


\subsection{Extracting a resetting sequence}



For an input sequence $c$ of length $l$ and the $n$ states of an FA, encoded in initial quantum states $\vert\mathcal{S}_{n-1}\rangle....\vert\mathcal{S}_{0}\rangle$, let $\bar{\vert\mathcal{S}}\rangle^{l,c}$ denote the collection (orederd list) of $n$ FA states, encoded in quantum states $\vert\mathcal{S}_{n-1}\rangle^l....\vert\mathcal{S}_{0}\rangle^l$  reached after the application of $c$ to the initial quantum states as indicated in equation eq.~\ref{eq:result3}. Similarly, for input sequence $c$, consider the related collection $\bar{\vert\mathcal{S}}\rangle^{l,c}$. These collections are considered to be the same (equivalent) if the following holds:


\begin{equation}
    \bar{\vert\mathcal{S}}\rangle^{l,c} = \bar{\vert\mathcal{S}}\rangle^{l,d} \implies (\vert\mathcal{S}_p\rangle^{l,c}=\vert\mathcal{S}_p\rangle^{l,d},\forall p=0...n-1).
    \label{eq:resss1}
\end{equation}

Then Eq.~\ref{eq:resss1} can factor eq.~\ref{eq:result3} as shown in eq.~\ref{eq:result4}. 
\begin{align}
\Delta^l\vert\bar{\mathcal{Q}_n}\rangle &=\vert\bar{\mathcal{Q}_n}\rangle^l\nonumber \\
&=\frac{1}{\sqrt{k^l}}\left (\vert\mathcal{X}^l_0\ldots\mathcal{X}^1_0\rangle\vert\mathcal{G}\rangle^{l,0}_{n-1}\ldots\vert\mathcal{G}\rangle^{1,0}_{n-1}\vert\mathcal{S}_{n-1}\rangle^{l,0}\ldots\vert\mathcal{G}\rangle_0^{l,0}\ldots\vert\mathcal{G}\rangle^{1,0}_{0}\vert\mathcal{S}_{0}\rangle^{l,0}
+\ldots+\vert\mathcal{X}^l_{i}\ldots\mathcal{X}^1_{h}\rangle\vert\mathcal{G}\rangle^{l,c}_{n-1}\ldots\vert\mathcal{G}\rangle^{1,c}_{n-1}\vert\mathcal{S}_{n-1}\rangle^{l,c}\nonumber  \right .\\&\left .\ldots\vert\mathcal{G}\rangle^{l,c}_{0}\ldots\vert\mathcal{G}\rangle^{1,c}_{0}\vert\mathcal{S}_{0}\rangle^{l,c}
+\ldots+\vert\mathcal{X}^l_{m}\ldots\mathcal{X}^1_{l}\rangle\vert\mathcal{G}\rangle^{l,d}_{n-1}\ldots\vert\mathcal{G}\rangle^{1,d}_{n-1}\vert\mathcal{S}_{n-1}\rangle^{l,d}\ldots\vert\mathcal{G}\rangle^{l,d}_{0}\ldots\vert\mathcal{G}\rangle^{1,d}_{0}\vert\mathcal{S}_{0}\rangle^{l,d}\nonumber \right .\\&\left .+\ldots+\vert\mathcal{X}^l_{k-1}\ldots\mathcal{X}^1_{k-1}\rangle\vert\mathcal{G}\rangle^{l,k^l-1}_{n-1}\ldots\vert\mathcal{G}\rangle^{1,k^l-1}_{n-1}\vert\mathcal{S}_{n-1}\rangle^{l,k^l-1}
\ldots\vert\mathcal{G}\rangle^{l,k^l-1}_{0}\ldots\vert\mathcal{G}\rangle^{1,k^l-1}_{0}\vert\mathcal{S}_{0}\rangle^{l,k^l-1} \right )\label{eq:result4}
\\&=\frac{1}{\sqrt{k^l}}\left (\vert\mathcal{X}^{1}_0\ldots\mathcal{X}^{1}_0\rangle\vert\mathcal{G}\rangle^{l,0}_{n-1}\ldots\vert\mathcal{G}\rangle^{1,0}_{0}\vert\bar{\mathcal{S}}\rangle^{l,0}
+\ldots+\tikzmark{M1}\left [\vert\mathcal{X}^l_{i}\ldots\mathcal{X}^1_{h}\rangle\vert\mathcal{G}\rangle^{l,c}_{n-1}\ldots\vert\mathcal{G}\rangle^{1,c}_{0}+\vert\mathcal{X}^l_{m}\ldots\mathcal{X}^1_{l}\rangle\vert\mathcal{G}\rangle^{l,d}_{n-1}\ldots\vert\mathcal{G}\rangle^{1,d}_{0}\right ]\tikzmark{M2}\tikzmark{M3}\vert\bar{\mathcal{S}}\rangle^{l,c}\tikzmark{M4}\nonumber \right .\\ &\left .+\ldots+\vert\mathcal{X}^l_{k-1}\ldots\mathcal{X}^1_{k-1}\rangle\vert\mathcal{G}\rangle^{k^l-1}_{n-1}\ldots\vert\mathcal{G}\rangle^{k^l-1}_{0}\vert\bar{\mathcal{S}}\rangle^{l,k^l-1} \right )\nonumber 
\label{eq:result4}
\end{align}

\begin{tikzpicture}[overlay,remember picture]
\draw[black, thick] (pic cs:M1)+(0,-.25) node[anchor=north,xshift=0.1cm,yshift=-0.05cm,xshift=3cm] (a1){B} rectangle ($(pic cs:M2)+(-0.05,.4)$);
\draw[black, thick] (pic cs:M3)+(0,-.25) 
rectangle ($(pic cs:M4)+(+0.1,.4)$)node[anchor=north,yshift=-0.7cm,xshift=-0.5cm] (a1){A};
\end{tikzpicture}

Equation~\ref{eq:result4} thus naturally factors into a form that includes both all input sequences with the associated garbage qubits (in Box B in eq.~\ref{eq:result4}) such that these sequences lead to the same collection of quantum states; here, $\vert\bar{\mathcal{S}}\rangle^{l,c}$ (in Box A in eq.~\ref{eq:result4}).


Using Grover algorithm on eq.~\ref{eq:result4}, with the oracle searching for the qubits representing a factored collection say $\bar{\vert\mathcal{S}}{\rangle^{l,c}}$ such that all quantum states of the collection correspond to the same FA state ; i.e., the following holds: eq.~\ref{eq:synch}. 
\begin{equation}
    \vert\mathcal{S}_p\rangle^{l,c}=  \vert\mathcal{S}_{p+1}\rangle^{l,c}, p=0..n-2
        \label{eq:synch}
\end{equation}


We apply the Grover algorithm~\cite{grover} with oracle searching for a collection where all its elements correspond to the same FA state. Finally upon measurement we identify a collection which is a solution and we obtain all input sequences (RSs) that reach that solution.
In general, in the worst-case, we reach $M=k^l$ collections; where the upper bound on $k^l$ is $n^n$. Let $M'$ be the number of collections with the same FA state, where the upper bound on $M'$ is $n$. Then the number of iterations required for the amplitude amplification to find a solution is $\sqrt{\frac{M}{M'}}$. Therefore, before applying the Grover algorithm, we first search for $M'$ using the Quantum Counting algorithm~\cite{Brassard_1998}. Then if $M'< (M/2)$, which is always the case as the $n$ is much smaller than $M/2$, we apply the Grover algorithm \cite{grover}.






The resulting state from eq.~\ref{eq:init85} is thus searched using the Grover algorithm on the qubits representing the quantum states. For instance, after three iterations of Grover algorithm, the quantum state from eq.~\ref{eq:init85} would results in the quantum state shown in eq.~\ref{eq:grovex}.

\begin{multline}
\sqrt{\frac{1}{4}}\left ( \vert 0000\rangle\vert 012\rangle +(\vert 0001\rangle+\vert 0100\rangle+\vert 0110\rangle+\vert 0111\rangle)\vert 001\rangle +(\vert 0010\rangle+\vert 0011\rangle)\vert 112\rangle+\vert 0101\rangle\vert 011\rangle \right . \\ \left .+(\vert 1000\rangle+\vert 1100\rangle+\vert 1110\rangle+\vert 1111\rangle)\vert 022\rangle +(\vert 1010\rangle+\vert 1011\rangle+\vert 1101\rangle)\vert 002\rangle\right )+\sqrt{\frac{3}{4}}\vert 1001\rangle\vert 222\rangle
\label{eq:grovex}
\end{multline}
\color{black}Such state would result with a probability of 75\% in measuring the the state $\vert222\rangle$ and thus indicating that the synchronizing sequence is $\vert1001\rangle$. The algorithm can be parameterized more precisely to give more dependable results after measurement. 


\section{Space complexity of finding a RS using our approach}


\subsection{Complexity}

The space complexity of the classical brute-force depth-first-search algorithm for an $n$-state FA for finding a resetting sequence is of polynomial order \textbf{$n\log(n)+n$}~\cite{eppstein}. In the presented quantum framework, for an $n$-state FA, the space complexity for finding a resetting sequence is linear.  A related proof is given in the supplementary material.



\begin{proof}

\begin{enumerate}
    \item First, $l\kappa$ applications of $H$ single qubit operator creates the initial input sequence superposition. 
    \item For each input $\vert\mathcal{X}\rangle^k$, from an input sequence of length $l$, there are $n$ applications of the $\Delta$ operator for a total of $l\times n$ operations.  
    \item In addition, each input value in the input sequence needs $C=2l^2\mu$ SWAP gates. First, let the input qubits be ordered by natural order $0\ldots \kappa-1$. Then requiring that the logical qubits must be spatially adjacent when $\Delta$ operator is applied, each qubit representing $\vert\mathcal{X}\rangle^k$ must be swapped with $2((k-1)\kappa)+2j(l\mu+\nu)$ qubits (two times). The first part $2((k-1)\kappa)$ is the number of Swap gates to move the input $\vert\mathcal{X}\rangle^k$ to the location of $\vert\mathcal{M}\rangle^1$. The second part $2j(l\mu+\nu)$, represents the number of SWAP gates required to move the input register $\vert\mathcal{X}\rangle^k$ close to the $\vert\mathcal{A}_0\rangle^{l}_{j}\ldots\vert\mathcal{A}_0\rangle^{1}_{j}\vert\mathcal{S}_{j}\rangle$ qubits, while crossing all $2(j-1)(l\mu+\nu)$ qubits. The largest term $2jl\mu$ is maximized by $2l^2\mu$ and thus is the largest bound for the number of SWAP gates. Note that the SWAP gates are required here only for the qubit routing and if the inputs qubits are repeated $n$-times, the number of SWAP gates is reduced to $2((k-1)\kappa)$.
    \item  Retrieval of one of the existing synchronizing sequence in $\sqrt{N}$ time. Once the simulation is finished, the final QTS $\vert\bar{\mathcal{Q}_n}\rangle^l$ contains a superposition of the $nl\mu$ and $l\kappa$ ancillae and input qubits respectively. It also contains a set of $n\nu$ qubits that contain $k^l$ possible collections of proportional to  $\vert\bar{\mathcal{S}}\rangle^{k,\sigma}$. However because these collections are not always different, the number is smaller and the upper  bound of required iterations of the Grover algorithm is thus upper bounded by $\sqrt{N}$ with $N=k^l$.
    \item Finally, $k^l>>2l^2\mu>n\times l$ and in addition considering the computational requirements only $n\times l$ is much smaller than $k^l$. As a result we consider that the most computationally intensive part remains the Grover algorithm with $\sqrt{k^l}$ steps. 
\end{enumerate}
In summary, the space complexity of the proposed method is linear in space requirements and quadratic in the time domain when compared to the brute-force search case. 
\end{proof}




For finding an RS, in a classical brute-force algorithm, one has to generate and evaluate all the $M=k^l$ input sequences in the worst case. Thus, the time complexity is exponentially given by $O(k^l)$. In the proposed method, the time complexity of simulating and generating all state combinations by applying the $l$ sequences to all states of the FA is of order $O(l)$. 

In addition, in a classical algorithm, each generated combination $x'$ is evaluated by a comparator. However, in the proposed quantum algorithm, the target combination needs to be searched using the amplitude amplification algorithm. The complexity of such a search is $\sqrt{\frac{M}{M'}}$. The solutions are the input sequences that lead to the same set of states as defined in Eq.~\ref {eq:synch}. When the number of target solutions is $1$, the amplitude amplification algorithm is equivalent to the Grover algorithm~\cite{grover}. Thus, the gain over a classical brute-force algorithm is quadratic. In addition, the speed of convergence of the amplitude amplification algorithm (Grover algorithm with multiple solutions) accelerates 
for $M'>1$. 

Note that while the classical algorithm generates $k^l$ collections of states, the quantum algorithm generates at most $n^n$ distinct collections. According to eq.~\ref{eq:result4}, collections with the same quantum states combine together. As a result, the amplitude amplification algorithm searches at a maximum of $n^n$ states. This also implies that while the gain of the proposed model is in the linear steps of generating all the collections of states $l$ versus $k^l$, the number of searched state collections is a relative gain $n^n$ versus $k^l$ when $k^l < n^n$. Moreover, when the quantum algorithm searches for $l: k^l<n^n$, the number of searched state collections is $k^l$.



\section{Conclusion}

 Finite state transition systems (FA) are used to represent hardware, software, mathematical, biological, and many other systems. We consider a well-known problem used in various domains which relates to finding an input sequence, called resetting sequence, which can bring the system to a known or safe state independent of its current state. In general the problem is exponential; and accordingly we solve the problem using a quantum model. To this end, we  use a proper representation of a given state machine in quantum space, and we search for an intended solution by exploring the behavior of the machine under all input sequences in superposition followed by a proper Grover search throughout the states reached by these sequences. 
A quadratic gain is provided by our approach over the exponential complexity of a traditional brute-force method, which is the only method that can be applied to a general FA class. 

The proposed approach paves the way for using the work in various related domains; and in particular, we focus now on adapting and extending the work in FSM-based testing \cite{lee96,dorofeeva2010}, where many types of input sequences, in addition to resetting sequences, are used in test derivation.

\bibliographystyle{plain}
\bibliography{Quantum}










\end{document}